\documentclass[%
preprintnumbers,
amsmath,amssymb,
aps,
]{revtex4-2}

\usepackage{graphicx}
\usepackage{dcolumn}
\usepackage{bm}
\usepackage{caption}
\usepackage{overpic}
\usepackage{color}
\usepackage[colorlinks, linkcolor=blue,
urlcolor=blue,
citecolor=blue]{hyperref}
\usepackage[mathlines]{lineno}
\usepackage{newtxtext}
\usepackage{newtxmath}

\begin{document}
	\preprint{}
	\title{A total-shear-stress-conserved wall model for large-eddy simulation of high-Reynolds number wall turbulence}
	\author{Huan-Cong Liu}
	\affiliation{AML, Department of Engineering Mechanics, Tsinghua University, 100084 Beijing, China}
	\author{Chun-Xiao Xu}
	\affiliation{AML, Department of Engineering Mechanics, Tsinghua University, 100084 Beijing, China}
	\author{Wei-Xi Huang}
	\email[E-mail: ]{hwx@tsinghua.edu.cn}
	\affiliation{AML, Department of Engineering Mechanics, Tsinghua University, 100084 Beijing, China}
	\date{\today}%
	
	\begin{abstract}
		Wall-modeled large-eddy simulation (WMLES) is widely recognized as a useful method for simulation of turbulent flows at high Reynolds numbers. Nevertheless, a continual issue in different wall models is the shift of the mean velocity profile from the wall-model/RANS (Reynolds-averaged Navier-Stokes) region to the LES region. This phenomenon, referred to as logarithmic layer mismatch (LLM), occurs in both wall shear stress models and hybrid RANS/LES models. Many efforts have been made to explain and resolve this mismatch, including decreasing the high correlation between the wall shear stress and the velocity at the matching layer, modifying the subgrid-scale (SGS) eddy viscosity, and adding a stochastic forcing. It is widely believed that the inclusion of the resolved Reynolds shear stress (or the convection term) is essential to elliminate the LLM, as it prevents the overseimation of the modeled Reynolds shear stress and promotes the generation of the small-scale flow structures in the near-wall region. In this work, by comparing three different SGS eddy viscosity models, we demonstrate that ensuring the total shear stress conservation (TSSC) conservation is key to resolving the LLM. Under the TSSC framework, the effect of the convection term on LLM can be quantitatively assessed. Furthermore, a modified SGS eddy viscosity modfication model that adheres to the TSSC constraint is tested at different Reynolds numbers ($Re_\tau=1000, 2000, 4200$). Our results demonstrate the robust performance of the present model in predicting skin friction and low-order turbulence statistics, even under a relatively low grid resolution ($\Delta x^+, \Delta z^+ \lesssim 500$, $2\leq \Delta_x/\Delta_{y,mat} \leq 4$, where $\Delta_{y,mat}$ is the wall-normal grid spacing in the wall-model region).
	\end{abstract}

	\maketitle
	
	\section{Introduction}\label{sec:introduction}
	   Unlike the Reynolds-averaged Navier-Stokes (RANS) method, which only captures the mean properties of turbulent flows, the large-eddy simulation (LES) technique has the ability to resolve energy-carrying eddies, making it well-suited for studying unsteady and multiscale turbulent flows. For wall turbulence, the use of wall-resolved LES (WRLES) is necessary to accurately depict all energy-carrying structures within the wall layer, leading to demanding resolution requirements at relatively high Reynolds numbers \citep{choi_grid-point_2012}. A cost-effective alternative solution is to employ a wall model or approximate boundary conditions to assist the coarse inner-layer mesh to represent the wall effects for the outer layer, known as wall-modeled LES (WMLES) \citep{piomelli_wall-layer_2002}.

	   WMLES can be generally classified into two categories \citep{larsson_large_2016,bose_wall-modeled_2018}: the hybrid RANS/LES formulation \citep{nikitin_approach_2000,piomelli_innerouter_2003,shur_hybrid_2008,spalart_comments_1997,spalart_new_2006}  and the wall-modeled formulation in which the LES governing equations are solved in the entire computational domain, while the wall model helps to introduce appropriate boundary conditions for the outer-layer flow \citep{bose_dynamic_2014,chen_reynolds-stress-constrained_2012,nicoud_large_2001,yang_integral_2015}. In the former category, the computational domain is divided into the near-wall region (RANS mode) and the remaining region (LES mode). The interface of these two regions can be explicitly prescribed \citep{piomelli_innerouter_2003} or implicitly designed by the closure model, as in detached-eddy simulation (DES) \citep{spalart_comments_1997}. In the latter category, the wall-stress model is the predominant and widely used method. Within this framework, information exchange occurs between the wall model and LES solver at both the wall and a certain distance from the wall (the location of the matching layer). The LES solver provides velocity information to the wall model at the matching layer, while the wall model supplies the wall shear stress to the LES solver at the wall. This exchange allows the wall-stress model to incorporate the effect of solid boundaries into the LES solver by replacing the no-slip boundary condition. This is particularly crucial when the grid resolution is too coarse to resolve sharp velocity gradients at the first grid away from the wall.

	   In the realm of wall models (including hybrid RANS/LES models and wall-stress models), a persistent challenge called the logarithmic layer mismatch (LLM), that is, the intercept of the velocity profile has an upward or downward shift from the upper boundary of the wall model region to the outer layer, has plagued the WMLES community for many years \citep{bose_wall-modeled_2018,cabot_near-wall_1996,hamba_hybrid_2003,kawai_dynamic_2013,piomelli_wall-layer_2002,piomelli_new_1989}. In the hybrid RANS/LES community, it was observed that the RANS model employed in the inner layer tends to suppress the formation of small-scale flow structures, inducing an unphysical buffer layer between the inner and outer layers \citep{piomelli_innerouter_2003,shur_hybrid_2008}. Various remedies have been proposed to encourage the generation of small-scale flow structures and thus ensure a smooth transition from the RANS region to the LES region. These include reducing the length scale \citep{shur_hybrid_2008}, modifying the subgrid-scale (SGS) eddy viscosity \citep{medic_formulation_2006,kalitzin_anear-wall_2007}, and introducing stochastic force in the near-wall region \citep{piomelli_innerouter_2003}. Additionally, Chen \textit{et al.} \cite{chen_reynolds-stress-constrained_2012} enforced a Reynolds stress constraint (RSC) on the SGS stress model in the inner layer, ensuring that the mean flow complies with the RANS solution, which successfully eliminated the LLM. In the community of wall shear stress models, it was shown that directly applying the velocity field from the grid adjacent to the wall model often leads to the underpredicted or overpredicted wall shear stress, ultimately resulting in the LLM. Some researchers attributed this issue to numerical and SGS modeling errors that inevitably contaminate the velocity field at the adjacent coarse grids \citep{kawai_dynamic_2013,nicoud_large_2001,wu_constraint_2013}. As a remedy, Kawai and Larrson \cite{kawai_wall-modeling_2012} suggested that the matching layer should be positioned further away from the wall than the adjacent grids. Yang \textit{et al.} \cite{yang_log-layer_2017} found that the velocity at the adjacent grids exhibits an unphysically high correlation with the calculated wall shear stress. Bou-Zeid \textit{et al.} \cite{bouzeid_largeeddy_2004} pointed out that the law of wall is validated in the ensemble-average sense, i.e., $\langle\tau_{w}\rangle\propto\langle u(y^{*})\rangle^{2}$. However, many algebraic wall models impose this law in the instantaneous and local sense, resulting in $\langle\tau_{w}\rangle\propto\langle u(y^{*})^{2}\rangle$, which causes an overestimation of the wall shear stress since $\langle u(y^{*})^{2}\rangle>\langle u(y^{*})\rangle^{2}$. Bou-Zeid \textit{et al.} \cite{bouzeid_largeeddy_2004} and Yang \textit{et al.} \cite{yang_log-layer_2017} both proposed to use spatially or temporally filtered LES data as input for the wall model, although they provided different explanations to this approach. Overall, the previous researches have given a comprehensive explanation of the physical mechanism behind the LLM and offered various remedies to successfully eliminate it. In the present study, we aim to provide an alternative explanation to the LLM in a more quantitative way.

	   In the canonical wall shear stress model, the calculated wall shear stress is only used to replace the viscous term at the wall, i.e., $\mu \frac{\partial U}{\partial y} \big|_{w} \equiv \tau_{w}$, in order to represent the sharp velocity gradients. However, the SGS stress models tend to underperform on coarse near-wall grids \citep{spalart_detached-eddy_2008,jimenez2000large,sayadi_large_2012,chen_reynolds-stress-constrained_2012}, leaving room for further improvement in simulation of high-Reynolds number wall turbulence. For this purpose, we explore the possibility of using the calculated wall shear stress to modify the SGS eddy viscosity within the wall-model region (see \ref{fig:schematic}). When only the SGS eddy viscosity at the wall is corrected (the eddy viscosity is non-zero at the wall), as detailed in $\S$\ref{sec:wall}, it is equivalent to replacing the no-slip boundary condition with the wall shear stress condition--an approach commonly utilized by numerous wall-stress models. Besides, it is worth noting that correcting the SGS eddy viscosity is not a new approach in the hybrid LES/RANS model, but it is less employed in the wall-stress model. Kalitzin \textit{et al.} \cite{kalitzin_anear-wall_2007} derived a RANS-like eddy viscosity corrected by the resolved Reynolds shear stress, with the wall shear stress obtained from a look-up table. Ma \textit{et al.} \cite{ma_hybrid_2021} derived a RANS-like eddy viscosity based on the constant shear stress approximation, which is more accurate than the widely used mixing-length model in the logarithmic layer and significantly reduces the resolution requirement of the first grid away from the wall. Chen \textit{et al.} \cite{chen_reynolds-stress-constrained_2012} enforced RSC on the SGS stress model, which is equivalent to correcting the SGS eddy viscosity based on the RANS results. 

	   In this work, by comparing three different subgrid-scale eddy-viscosity-modification (SGS-EVM) models, we provide a new explanation to the LLM phenomenon and highlight the importance of non-equilibrium effects (the pressure gradient effect and the convection effect) even in the equilibrium flows. While the previous studies have shown that incorporating non-equilibrium effects can improve simulation results and alleviate the LLM\citep{wang_dynamic_2002,kalitzin_anear-wall_2007,kawai_dynamic_2013,park_improved_2014,chen_reynolds-stress-constrained_2012}, our focus is on providing a more quantitative understanding of the relationship between the non-equilibrium effects and the LLM.  
	   The remainder of this paper is organized as follows. In $\S$\ref{sec:computational}, the governing equations and numerical method are introduced, including the formulation of the total shear-stress conservation (TSSC) model. In $\S$\ref{sec:positive}, we examine the formation of positive LLM and discuss the effects of pressure gradient and convection terms on the LLM phenomenon. At the end of $\S$\ref{sec:positive}, we revisit the physical mechanism of the TSSC model and provide detailed comparisons with the previous remedies for LLM. In $\S$\ref{sec:TSSC}, we demonstrate the efficiency and robustness of the TSSC model in high-Reynolds number turbulent channel flow. Finally, conclusions are drawn in $\S$\ref{sec:conclusions}.

	\section{Computational methods}\label{sec:computational}

		\subsection{LES solver}\label{sec:LES}
			For Newtonian incompressible flows, the following low-pass filtered Navier-Stokes (N-S) and continuity equations are solved:
			\begin{equation}\label{eq:LES}
				\begin{aligned}
					\frac{\partial\bar{u}_{i}}{\partial t}+\frac{\partial\bar{u}_{i}\bar{u}_{j}}{\partial x_{j}}&=-\frac{\partial\bar{p}^*}{\partial x_{i}}+\nu\frac{\partial^{2}\bar{u}_{i}}{\partial x_{j}\partial x_{j}}-\frac{\partial\bar{\tau}_{ij}^{LES,d}}{\partial x_{j}} , \\\frac{\partial\bar{u}_{i}}{\partial x_{i}}&=0 .
				\end{aligned}
			\end{equation}
			
            \noindent In the above equations, $(x_1,x_2,x_3)$ or equivalently $(x,y,z)$ denote the streamwise, wall-normal and spanwise axes, respectively; $(u_1,u_2,u_3)$ or $(u,v,w)$ are the corresponding velocity components; $\bar{p}^{*}=\frac{\bar{p}}{\rho}+\frac{1}{3}\bar{\tau}_{kk}^{LES}$, where $p$ is the pressure and $\rho$ is the density; $\nu$ is the kinematic viscosity of fluid. Here, the overbar denotes low-pass filtering with a filter width $\Delta$ defined as the cube root of the grid volume, i.e. $\Delta=\sqrt[3]{\Delta_{x}\Delta_{y}\Delta_{z}}$.  The stress term $\bar{\tau}_{ij}^{LES}=\overline{u_{i}u_{j}}-\bar{u}_{i}\bar{u}_{j}$ represents the effects of the SGS flow on the resolved scales, and here we use the dynamic Smagorinsky model \citep{germano_dynamic_1991,lilly_proposed_1992} for the anisotropic part of ${\bar{\tau}}_{ij}^{LES}$:
			\begin{equation}\label{eq:tauij}
				\bar{\tau}_{ij}^{LES,d}=\bar{\tau}_{ij}^{LES}-\frac{1}{3}\delta_{ij}\bar{\tau}_{kk}^{LES}=-2\bar{\nu}_{sgs}\bar{S}_{ij}=-2C_{s}^{2}\Delta^{2}\left|\bar{S}\right|\bar{S}_{ij},
			\end{equation}
			where ${\bar{\nu}}_{sgs}$ denotes the SGS eddy viscosity, ${\bar{S}}_{ij}$ denotes the resolved-scale strain-rate tensor and $|\bar{S}|=\left(2\bar{S}_{ij}\bar{S}_{ij}\right)^{1/2}$. The model coefficient $C_s$ is assumed to be scale-invariant and is determined by the dynamic procedure, where a subtest filter with a width of $2\Delta$ is adopted. In simulation, the total viscosity (summation of $\nu$ and ${\bar{\nu}_{sgs}}$) is set as zero when its primary value is negative. Furthermore, different values of  ${\bar{\nu}}_{sgs}$  are used in the streamwise, wall-normal, and spanwise momentum equations, denoted as ${\bar{\nu}}_{sgs}^x$, ${\bar{\nu}}_{sgs}^y$ and ${\bar{\nu}}_{sgs}^z$, respectively. While this may seem to break the symmetry of the subgrid-scale stress tensor, it is justified because the calculated SGS eddy viscosity not only represents the subgrid-scale stress but also compensates for discretization errors in the velocity gradient on the coarse grid. Due to the anisotropic nature of these discretization errors, it is necessary for the SGS eddy viscosity to vary across directions; otherwise, the velocity gradient would need to be corrected. This is why the SGS eddy viscosity at the wall is non-zero, to compensate for the velocity gradient (see \ref{sec:wall}).



			For spatial discretization, the second-order central finite-difference scheme is applied on a staggered Eulerian grid. The scalar variables including pressure are stored at the cell center and the velocities are stored at the centers of cell faces.  The SGS stress term is discretized in a conserved form, and the eddy viscosity is calculated at the cell faces by interpolating from the values at the cell centers. For temporal discretization, the flow solver is advanced by the fully implicit Crank-Nicholson scheme. The fractional step algorithm in conjunction with the matrix decomposition scheme is adopted for velocity-pressure decoupling and velocity-component decoupling. Readers can refer to Kim \textit{et al.} \cite{kim_implicit_2002} for more details.
			
		\subsection{Wall model}\label{sec:wall}
			\begin{figure}
				\centering
				\includegraphics[width=0.8\linewidth]{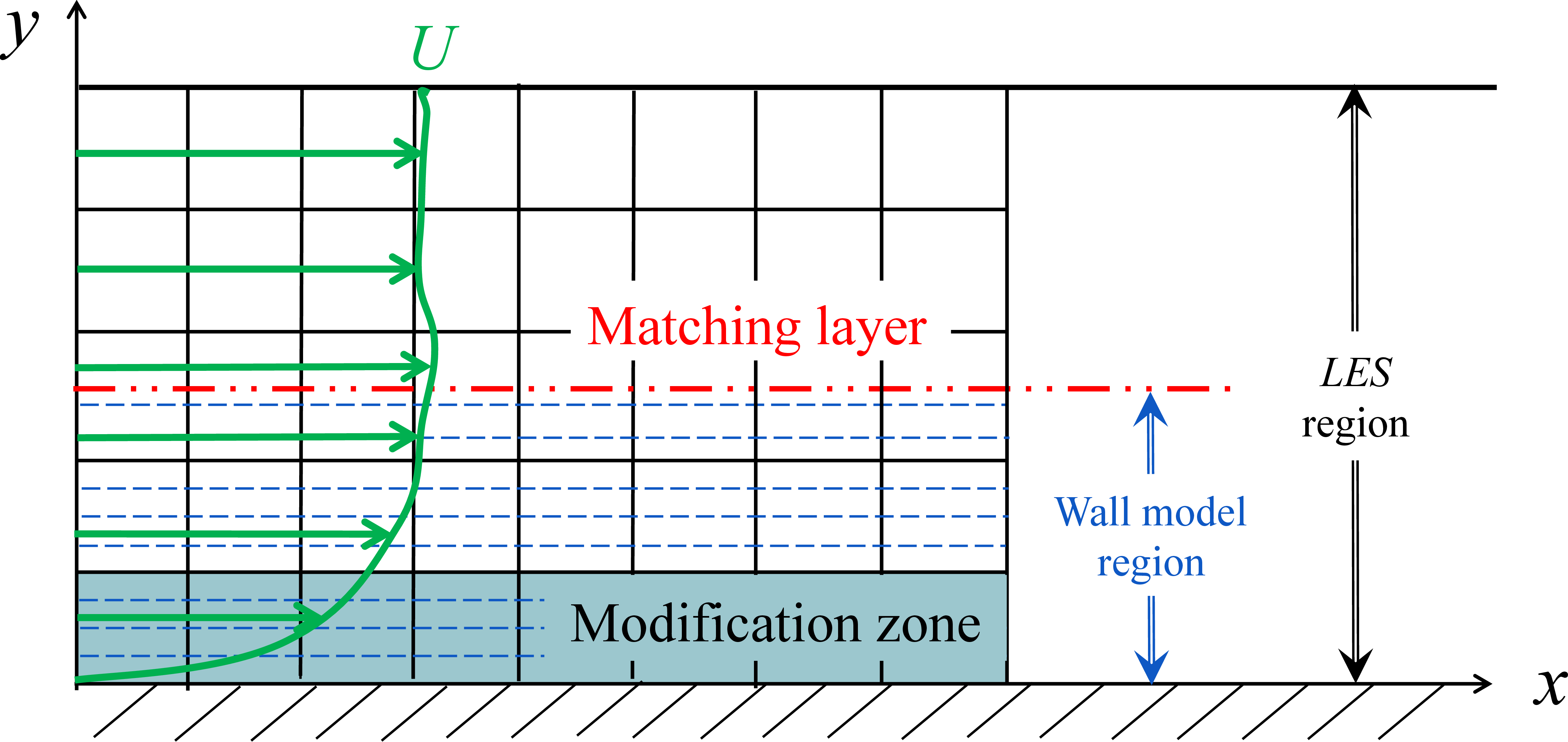}
				\captionsetup{justification=raggedright,singlelinecheck=true}
                \caption{Schematic of wall model implementation. The black solid lines represent the mesh for the LES solver. The blue dashed lines refined in the wall-normal direction represent the mesh for solving the wall model.}
				\label{fig:schematic}
			\end{figure}
			In the present study, $\bar{\nu}_{sgs}^x$ is corrected by wall model to maintain the conservation of total shear stress. $\bar{\nu}_{sgs}^y$ and $\bar{\nu}_{sgs}^z$ can be also corrected using alternative formulations, for simplicity, they are calculated using the dynamic Smagorinsky model. The implementation of the present wall model is depicted in figure \ref{fig:schematic}. In the wall-model region, the streamwise thin boundary layer equation (sTBLE, see Eq.(\ref{eq:sTBLE})) is solved by a separated refined mesh to obtain the local wall shear stress. Then, the calculated $\tau_w$ is substituted in the SGS-EVM model and the modified $\bar{\nu}_{sgs}^x$ within the modification zone (as marked by the green region in figure \ref{fig:schematic}) can be obtained. In the present study, the matching layer is set at the center of the third grid, and the modification zone can cover 0 to 3 layers of grids away from the wall. Specifically, 'cover zero layer of grid' means that only the eddy viscosity at the wall is corrected by the SGS-EVM model. According to the introduction in §\ref{sec:introduction}, the present wall model is classified as a wall-stress model. Here, the calculated wall shear stress is used not only to replace the no-slip boundary condition, but also to correct $\bar{\nu}_{sgs}^x$ in the modification zone. For simplicity, $\bar{\nu}_{sgs}^x$ will be written as $\bar{\nu}_{sgs}$ hereafter.
			
			The sTBLE is expressed as:
			\begin{equation}\label{eq:sTBLE}
				\frac{\partial}{\partial y}\biggl[(\nu+\nu_{t})\frac{\partial u}{\partial y}\biggr]=\frac{\partial u}{\partial t}+\frac{\partial uu}{\partial x}+\frac{\partial u\nu}{\partial y}+\frac{1}{\rho}\frac{\partial p}{\partial x},
			\end{equation}
			where ${\nu_t}$ is the RANS-type eddy viscosity and can be calculated by the mixing-length model with near-wall damping \citep{van_driest_turbulent_1956, cabot_approximate_2000}, i.e., 
			\begin{equation}\label{eq:mixing-length}
				\begin{gathered}
					\nu_{t}=\kappa u_{\tau}yD(y)=\nu\cdot\kappa y^{+}D(y^{+}), \\
					D(y)=\left(1-e^{-y^{+}/A}\right)^{2}.
				\end{gathered}
			\end{equation}
			Here the superscript '+' denotes the normalization based on the wall viscous unit; $\kappa=0.41$ is the von Kármán constant and $A=19$.
			
			At each time step, Eqs. (\ref{eq:sTBLE}) and (\ref{eq:mixing-length}) are solved iteratively to obtain $\tau_w$ within the wall-model region with a refined one-dimensional mesh. The upper boundary condition is the streamwise velocity from the LES solver, while the no-slip boundary condition is applied at the wall. The RANS-type eddy viscosity ($\nu_t$) should be adjusted to match ${\bar{\nu}}_{sgs}$ at the matching layer if Eq. (\ref{eq:sTBLE}) contains the nonlinear convection term. On the other hand, if the unsteady term and the convection term are omitted, then $\kappa$ should remain constant in the fully developed turbulent channel flow \citep{kawai_dynamic_2013,park_improved_2014,wang_dynamic_2002}. For simplicity, here we adopt the second strategy.
			
			Numerous studies have used the thin boundary layer equation to calculate the wall shear stress ($\tau_w$) and improvements have been proposed for better accuracy  \citep{balaras_subgrid-scale_1994,hoffmann_approximate_1995,ma_dynamic_2019,ma_hybrid_2021,wang_dynamic_2002}. On the other hand, in the present study we focus on how to feed the obtained $\tau_w$ back to the LES solver more appropriately. We declare that to eliminate LLM, the appropriate feedback is an indispensable requirement for a wall model, which involves adjustment of ${\bar{\nu}}_{sgs}$ in the modification zone. A brief derivation of the SGS-EVM model is introduced as follows.	
   
			By integrating the filtered N-S equations in the $y$-direction, we obtain:
			\begin{equation}
				\int_{0}^{y}\Bigg(\frac{\partial\bar{u}_{i}}{\partial t}+\frac{\partial\bar{u}_{i}\bar{u}_{j}}{\partial x_{j}}\Bigg)dy=\int_{0}^{y}\Bigg(-\frac{\partial\bar{p}^{*}}{\partial x_{i}}+\nu\frac{\partial^{2}\bar{u}_{i}}{\partial x_{j}\partial x_{j}}-\frac{\partial\bar{\tau}_{ij}^{LES,d}}{\partial x_{j}}\Bigg)dy.
			\end{equation}
			After performing the ensemble average on both sides of the above equation, it is reasonable to omit the unsteady term, and the streamwise and spanwise gradient terms for the fully developed turbulent channel flow. In addition, the eddy-viscosity model is used for the SGS stress, i.e., $\bar{\tau}_{ij}^{LES,d}=-2\bar{\nu}_{sgs}\bar{S}_{ij}$. Finally, we can get the simplified momentum equation in the $x$-direction:
			\begin{equation}\label{eq:simplified-eq-x}
				\left\langle\bar{u}\bar{v}\right\rangle=-\frac{\left\langle\tau_{w}\right\rangle}{\rho}-\frac{\partial\left\langle\bar{p}^{*}\right\rangle}{\partial x}y+\nu\frac{\partial\left\langle\bar{u}\right\rangle}{\partial y}+\left\langle\bar{\nu}_{sgs}\frac{\partial\bar{u}}{\partial y}\right\rangle+\left\langle\bar{\nu}_{sgs}\frac{\partial\bar{\nu}}{\partial x}\right\rangle.
			\end{equation}
			Here the angular bracket represents the ensemble average, including the streamwise, spanwise and temporal average. By rearranging the above equation, and ignoring the relatively small term $\left\langle\bar{\nu}_{sgs}\frac{\partial\bar{v}}{\partial x}\right\rangle$ and the fluctuation product term $\left\langle\bar{\nu}_{sgs}^\prime \frac{\partial\bar{u}^\prime}{\partial y}\right\rangle$, we obtain the final SGS-EVM model:
			\begin{equation}\label{eq:TSSC-equation}
				\left\langle\bar{\nu}_{sgs}\right\rangle \approx 
                    \left[\frac{\left\langle\tau_{w}\right\rangle}{\rho}+\frac{\partial\left\langle\bar{p}^{*}\right\rangle}{\partial x}y+\left\langle\bar{u}\bar{v}\right\rangle\right]\left/\frac{\partial\left\langle\bar{u}\right\rangle}{\partial y}-\nu.\right. 
			\end{equation}
			Only $\langle\tau_{w}\rangle $ in the above equation should be calculated by the wall model, and other terms on the right hand side (including the velocity gradient term) can be directly obtained from the LES solver. It is worth mentioning that the above SGS-EVM model is directly derived from the filtered streamwise momentum equation, and no assumption is made except that some relatively small terms are neglected and some terms are removed for the specific flows (such as the fully developed channel flow). In addition, only the ensemble average of the SGS eddy viscosity is calculated, and from the a posteriori results (see $\S$\ref{sec:TSSC}), it is acceptable to replace the actual eddy viscosity with its ensemble avarage value for a correct prediction of low-order turbulence statistics.
			
			For the sTBLE equation (\ref{eq:sTBLE}), the previous studies systematically revealed the effects of the pressure gradient term and the convection term \citep{duprat_wall-layer_2011,hoffmann_approximate_1995,larsson_large_2016,wang_dynamic_2002}. In the equilibrium state, it is a prevailing assumption that the right-hand side of Eq.(\ref{eq:sTBLE}), which accounts for unsteadiness, convection, and pressure gradient effects, can be disregarded. This is primarily due to the facts that the LES time step is much larger than the characteristic timescale of the near-wall eddies, and the flow remains uninfluenced by external non-equilibrium effects. Even in the non-equilibrium state, Larsson \textit{et al.} \cite{larsson_large_2016} pointed out that the pressure gradient term and the convection term can also be neglected together because they almost balance each other. As mentioned above, the implementation of wall model consists of two main steps. The first step is to accurately determine the wall shear stress, and the second step is to appropriately reintroduce the calculated wall shear stress to the LES solver. While it is acceptable to ignore the non-equilibrium effects when solving the sTBLE in the equilibrium flow, it is not obvious that the non-equilibrium effects can also be disregarded when adjusting the eddy viscosity in the modification zone. Actually, resolved Reynolds shear stress has been proved to be crucial for the accuracy of the simulation results \citep{wang_dynamic_2002,kalitzin_anear-wall_2007,kawai_dynamic_2013,park_improved_2014,chen_reynolds-stress-constrained_2012}. In the following section, we attempt to explain the non-equilibrium effects on the LLM phenomenon in a more quantitative manner.
			
			In the next section, we will compare the three different SGS-EVM models, as listed in table~\ref{tab:SGS-EVM models}, to better elucidate the essential role of the convection term in Eq.(\ref{eq:TSSC-equation}). Specifically, the model associated with Eq.(\ref{eq:TSSC-equation}) will be referred to as the WMpcn model or the total-shear-stress-conserved (TSSC) model in the subsequent discussion. The capitalized 'WM' denotes the abbreviation of 'wall model', and the subscripts of 'p' and 'c' represent the pressure gradient term and convection term, respectively. In addition, the subscript 'n' denotes the number of layers in the modification zone. Thus, the WMn model neglects these two terms, while the WMpn model neglect the convection term. For all the three models, when the layer of grid $n=0$, the eddy viscosity at the wall satisfies $\bar{\nu}_{\mathrm{sgs}}=\frac{\left\langle\tau_{w}\right\rangle}{\rho}\left/\frac{\partial\left\langle\bar{u}\right\rangle}{\partial y}-\nu\right.$, which is equivalent to replacing the viscous term with the calculated wall shear stress, as is commonly done in most wall-stress models \citep{kawai_wall-modeling_2012,lee_large_2013}. Obviously, the SGS eddy viscosity at the wall is non-zero to compensate for the velocity gradient.
		
		\begin{table}
			\captionsetup{justification=raggedright,singlelinecheck=true}
			\caption{Formulations of the three SGS-EVM models. The capitalized 'WM' denotes the abbreviation of 'wall model'. The lowercase letters 'p' and 'c' represent the pressure term and the convection term, respectively. The appearance of 'p' or 'c' indicates that the corresponding term is included in the model. The lowercase letter 'n' means that the modification zone covers 'n' layers of grids, with $n=0,1,2,3$ in this paper.}
			\label{tab:SGS-EVM models}
			\begin{ruledtabular}
				\begin{tabular}{ll}
					\textrm{Name}&
					\textrm{SGS-EVM model} \\
					\colrule
					WMn  &  $\bar{\nu}_{sgs}=\frac{\left\langle\tau_w\right\rangle}{\rho}\left/\frac{\partial\left\langle\bar{u}\right\rangle}{\partial y}-\nu\right.$ \\
					WMpn &  $\bar{\nu}_{sgs}=\left[\frac{\left\langle\tau_w\right\rangle}{\rho}+\frac{\partial\left\langle\bar{p}^*\right\rangle}{\partial x}y\right]\left/\frac{\partial\left\langle\bar{u}\right\rangle}{\partial y}\right.-\nu$ \\
					WMpcn &  $\bar{\nu}_{sgs}=\left[\frac{\left\langle\tau_w\right\rangle}\rho+\frac{\partial\left\langle\bar{p}^*\right\rangle}{\partial x}y+\langle\bar{u}\bar{v}\rangle\right]\left/\frac{\partial\left\langle\bar{u}\right\rangle}{\partial y}\right.-\nu $ \\
				\end{tabular}
			\end{ruledtabular}
		\end{table}
			
	\section{Positive LLM}\label{sec:positive}
	   In this section, we study the LLM and compare different SGS-EVM models in the turbulent channel flow at $Re_{\tau}=550$. The computational domain is $2\pi\delta\times2\delta\times\pi\delta$ in the streamwise, normal and spanwise directions, respectively, with $\delta$ the half channel height. The grid numbers are $32\times30\times32$ with uniform grid along the streamwise and spanwise directions ($\mathrm{\Delta}_x^+\approx108$, $\mathrm{\Delta}_z^+\approx54$) and stretched grid along the normal direction ($\mathrm{\Delta}_y^+\approx30-43$). Moreover, we adopt 64 uniform grids inside the wall-model region for solving the sTBLE.
	
	   In the following, we will give an explanation of the positive LLM by using the WMn model (see  $\S$\ref{tab:SGS-EVM models}). Then, the effects of pressure gradient and convection on the LLM phenomenon will be discussed in $\S$\ref{sec:Pressure} and $\S$\ref{sec:convection}, respectively. In $\S$\ref{sec:revisiting}, some prevailing explanations and remedies of LLM will be revisited, together with the proposed solution.

	\subsection{Formation of the positive LLM}\label{sec:formation}
	
	\begin{table}
		\captionsetup{justification=raggedright,singlelinecheck=true}
		\caption{Comparison of wall shear stress from the wall model ($\tau_{w,WM}$), the LES solver ($\tau_{w,WMLES}$), and the DNS data ($\tau_{w,DNS}$) at ${Re}_\tau=550$ \cite{lee_direct_2015} with the WMn ($n=0-3$) models. Here,  $\varepsilon_{LW}=\left(\tau_{w,WMLES}-\tau_{w,WM}\right)/\tau_{w,WM}\times100\%$, $\varepsilon_{LD}=\left(\tau_{w,WMLES}-\tau_{w,DNS}\right)/\tau_{w,DNS}\times100\%$, with $\tau_{w,WMLES}=\langle-\frac{\partial\bar{p}}{\partial x}\delta\rangle$.}
		\label{tab:WMn-models}
		\begin{ruledtabular}
			\begin{tabular}{lcccc}
				\textrm{Case}&
				\textrm{WM0}&
				\textrm{WM1}&
				\textrm{WM2}&
				\textrm{WM3} \\
				\colrule
				$\varepsilon_{LW}(\%)$ & -0.08 & 0.28 & -0.13 & 0.01 \\
				$\varepsilon_{LD}(\%)$ & -1.49 & -1.69 & -3.68 & -3.50 \\
			\end{tabular}
		\end{ruledtabular}
	\end{table}
	
	\begin{figure}
		\centering
		\begin{overpic}[width=0.48\linewidth
			]{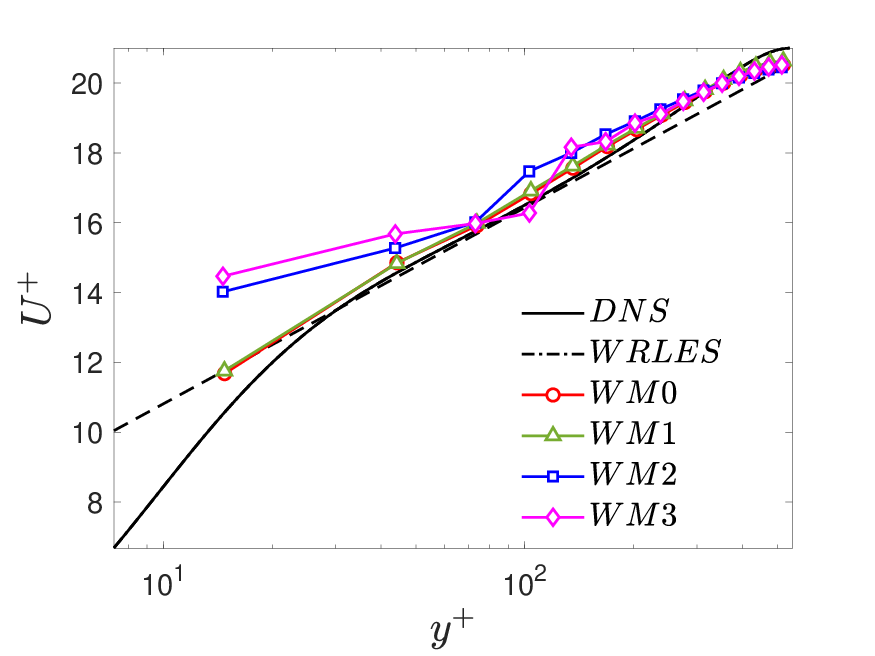}
			\put(7,73){($a$)}
			\put(30,60){\color{red}LLM region}
			\put(61,52){\color{red}\circle{20}}
		\end{overpic}
		\begin{overpic}[width=0.48\linewidth
			]{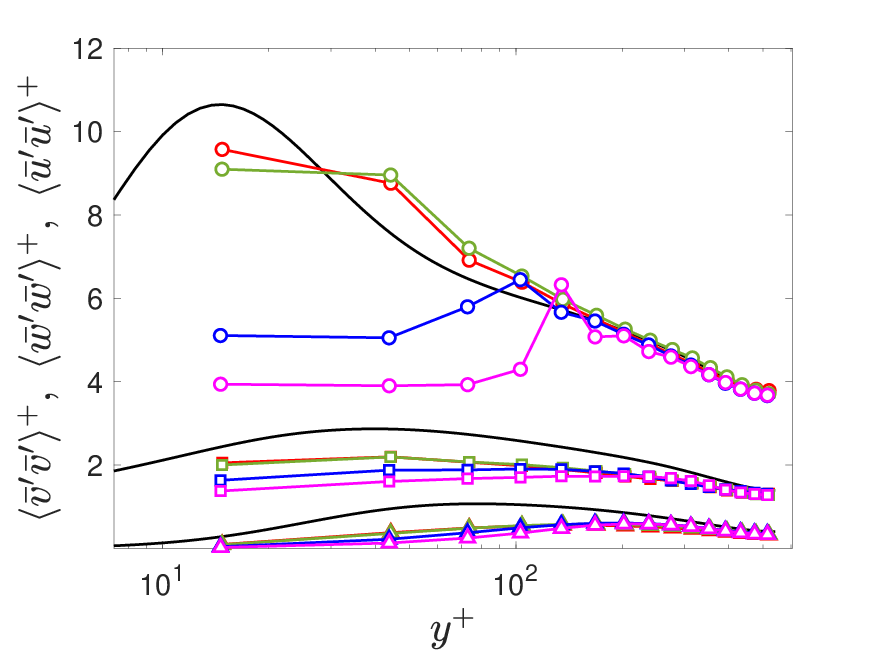}
			\put(7,73){($b$)}
		\end{overpic}
		\begin{overpic}[width=0.48\linewidth
			]{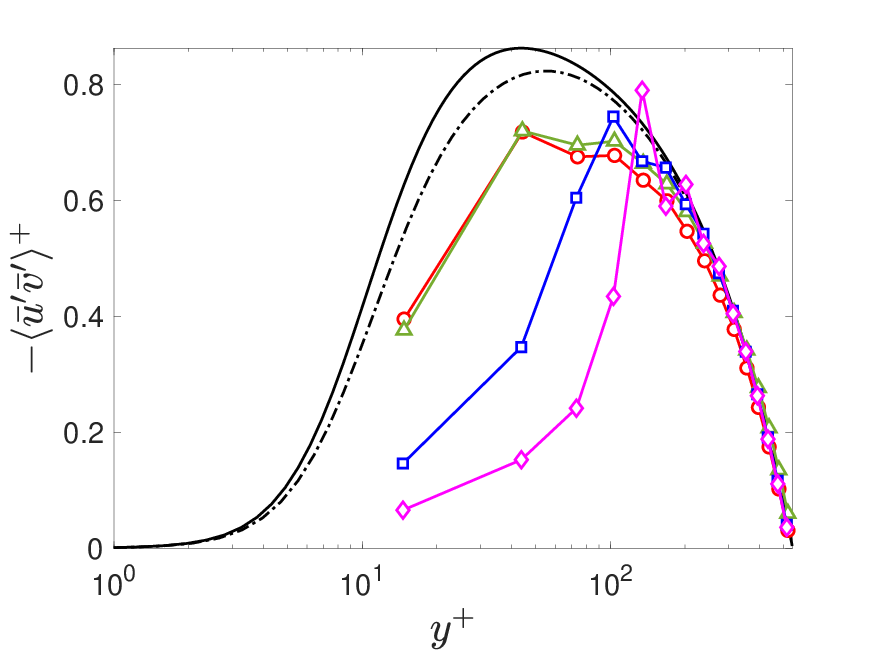}
			\put(7,73){($c$)}
		\end{overpic}
		\begin{overpic}[width=0.48\linewidth
			]{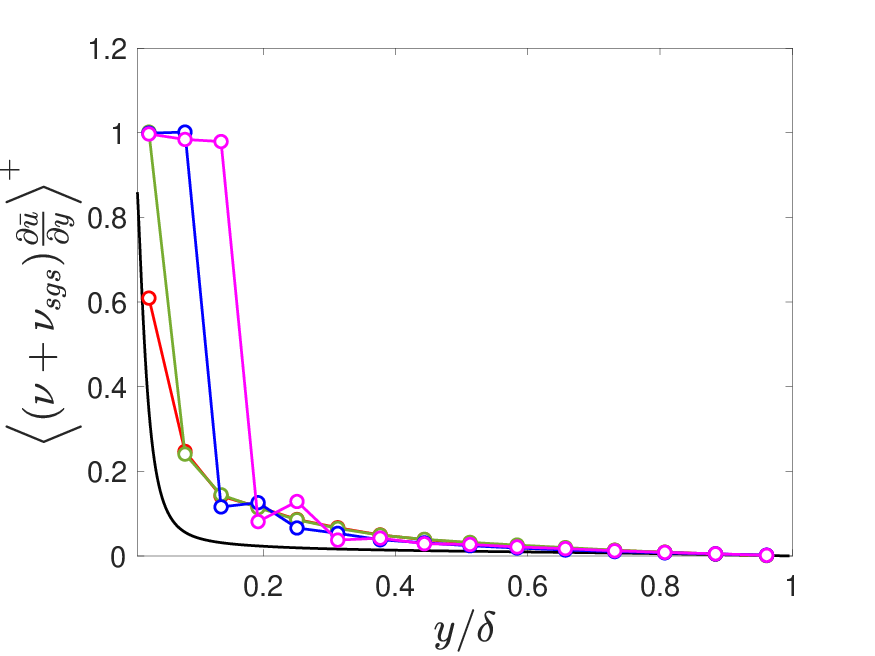}
			\put(7,73){($d$)}
		\end{overpic}
		\begin{overpic}[width=0.48\linewidth
			]{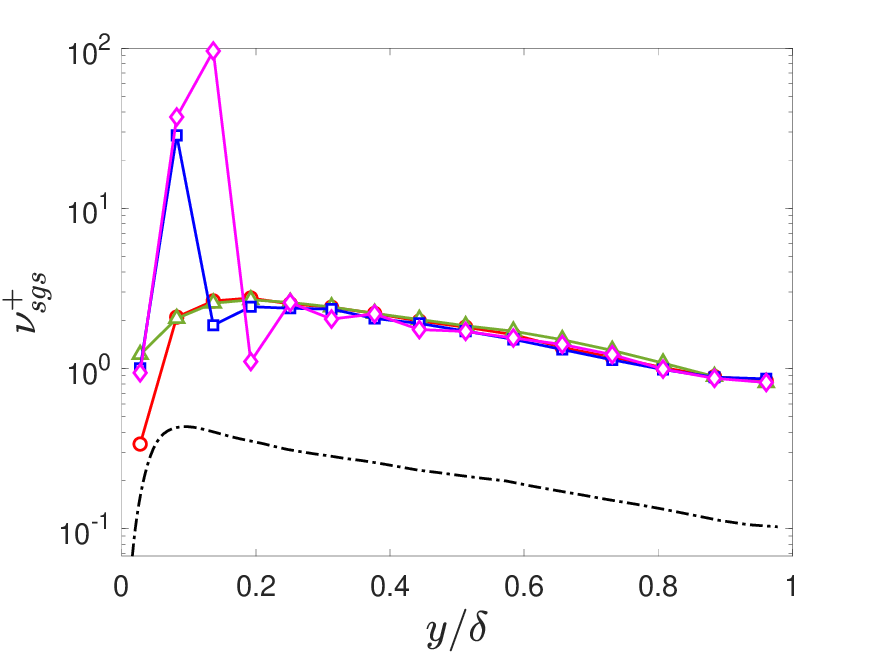}
			\put(7,73){($e$)}
		\end{overpic}
		\begin{overpic}[width=0.48\linewidth
			]{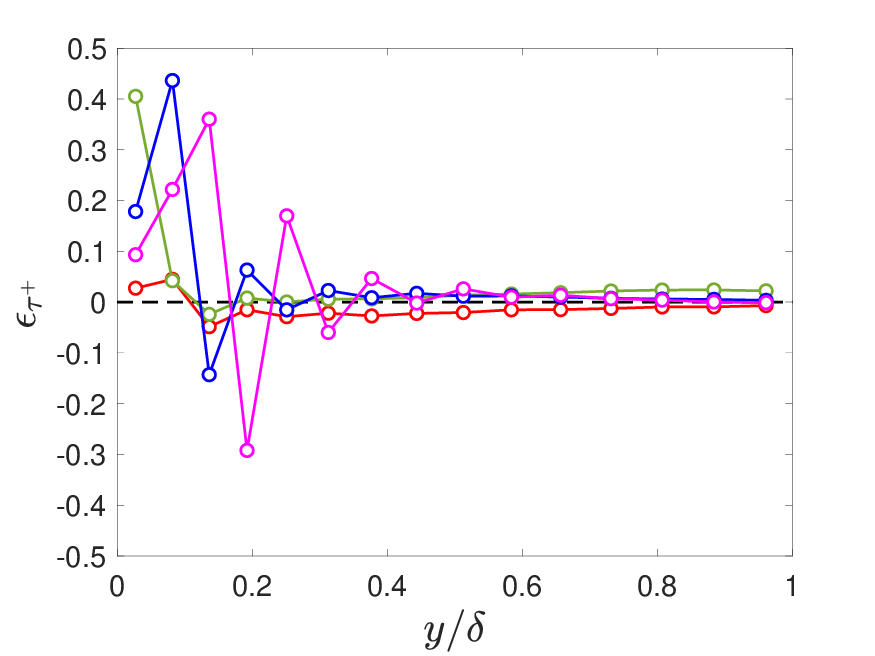}
			\put(7,73){($f$)}
		\end{overpic}\\
		\quad\\
		
		\captionsetup{justification=raggedright,singlelinecheck=true}
		\caption{(Colour online) Simulation results from the WMn ($n=0-3$) models at ${Re}_\tau=550$: (a) profiles of mean velocity $U=\langle \bar{u} \rangle$, where the black dashed line denotes the logarithmic law, i.e. $U^+=\frac{1}{0.41}ln(y^+)+5.2$; (b) the resolved Reynolds normal stresses, where the streamwise and spanwise components are shifted upward by 3 units and 1 unit, respectively, for clarity; (c) the resolved Reynolds shear stress; (d) summation of the viscous stress and the modeled Reynolds shear stress; (e) the SGS eddy viscosity; (f) deviation of the total shear stress, i.e. ${\epsilon_{{\tau ^ + }}} = {\left\langle \rho {\left( {\nu  + {{\bar \nu }_{sgs}}} \right)\frac{{\partial \bar u}}{{\partial y}}} \right\rangle ^ + } - {\left\langle {\rho \bar u\;\bar v} \right\rangle ^ + } - \left( {1 - \frac{y}{\delta }} \right)$. The DNS data from Lee and Moser \cite{lee_direct_2015} and the WRLES result are also plotted for comparison, where the WRLES is performed by the flow solver introduced in $\S$\ref{sec:LES} and the grid numbers are $128\times128\times128$, with uniform grids along the horizontal directions and stretched grids along the wall-normal direction. The superscript '+' denotes the normalization in wall units.}  
		\label{fig:WMn-models}               
	\end{figure}

	\begin{figure}
		\centering
		\includegraphics[width=0.6\linewidth]{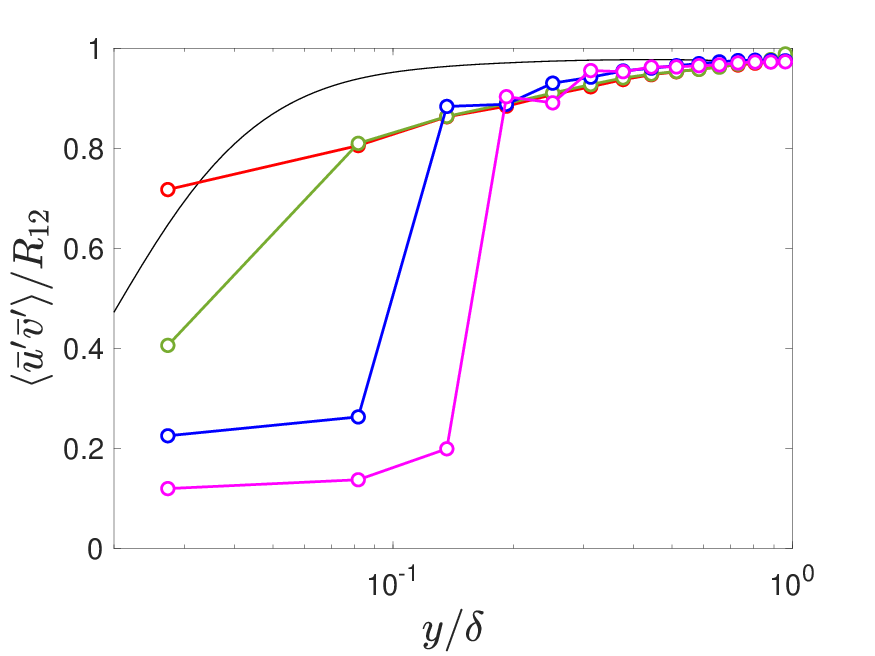}
		\captionsetup{justification=raggedright,singlelinecheck=true}
		\caption{\label{fig:WMn-ratio}
			(Colour online) The ratio of the resolved Reynolds shear stress to the total shear stress ($R_{12}$) for the WMn ($n=0-3$) models at $Re_{\tau}=550$, where $R_{12}=\langle -\bar{u}^\prime \bar{v}^\prime + \nu_{sgs}\frac{\partial\bar{u}}{\partial\bar{y}} \rangle$.}
	\end{figure}
	
	Table \ref{tab:WMn-models} shows a comparison of wall shear stress from wall model (${\tau _{w,WM}}$), LES solver (${\tau _{w,WMLES}} = \left\langle { - \frac{{\partial {\bar p}^*}}{{\partial x}}\delta } \right\rangle $) and DNS data (${\tau _{w,DNS}}$). The deviation of ${\tau _{w,WMLES}}$ from ${\tau _{w,WM}}$ is define as $\varepsilon_{LW}=(\tau_{w,WMLES}-\tau_{w,WM})/\tau_{w,WM}\times 100\%$. Moreover, the deviation of ${\tau _{w,WMLES}}$ from ${\tau _{w,DNS}}$ is defined as $\varepsilon_{LD}=(\tau_{w,WMLES}-\tau_{w,DNS})/\tau_{w,DNS}\times 100\%$. It can be seen in table \ref{tab:WMn-models} that all the $\left| {{\varepsilon _{LW}}} \right|$ are within 0.5$\%$, which means that the wall model adequately transfer the wall shear stress to the flow field in a mean sense; all the $\left| {{\varepsilon _{LD}}} \right|$ are within 4$\%$, which are much smaller than the observed 10$\%$-15$\%$ wall friction error when the LLM phenomenon occurs as reported by Larsson \textit{et al.} \cite{larsson_large_2016}. Unfortunately,  for the cases of WM2 and WM3, an obvious positive LLM can be still observed in figure \ref{fig:WMn-models}(a) with a low slope in the near-wall region and an abrupt increase in slope beyond the modification zone. In other words, small ${\varepsilon_{LW}}$ and ${\varepsilon_{LD}}$ are not sufficient conditions to avoid the LLM phenomenon.
	
	For turbulent channel flow, the total shear stress is expected to distribute linearly along the wall-normal direction, i.e., ${\tau ^ + } = {\left\langle {\left( {\nu  + {{\bar \nu }_{sgs}}} \right)\frac{{\partial \bar u}}{{\partial y}}} \right\rangle ^ + } - {\left\langle {\bar u\;\bar v} \right\rangle ^ + } = \left( {1 - \frac{y}{\delta }} \right)$. Figure \ref{fig:WMn-models}(f) shows the deviation of the total shear stress from the theoretical value. When the number of modification layer is zero (the WM0 case) or one (the WM1 case), the mean velocity profile (figure \ref{fig:WMn-models}a) collapses well with the DNS data, and the resolved Reynolds stresses (figure \ref{fig:WMn-models}b,c) all exhibit a smooth distribution. It is reasonable that the resolved Reynolds shear stress is lower than the DNS or WRLES result, because the grid is too coarse to resolve all the energy-containing flow structures. With increased coverage of the modification zone, the WMn model overestimates the SGS eddy viscosity (see figure \ref{fig:WMn-models}e) and the summation of the modeled Reynolds shear stress and the viscous stress nearly equals to the wall shear stress within the modification zone (see figure \ref{fig:WMn-models}d). This leads to excessive total shear stress inside the modification zone and significant suppression of the near-wall fluctuations (see figure \ref{fig:WMn-models}b,c). Moreover, the turbulent intensities and resolved Reynolds shear stress are weakened not only in the modification zone but also in the layers beyond that.  Outside the modification zone, the ${\bar \nu _{sgs}}$ calculated by the dynamic Smagorinsky model is not enough to supply sufficient modeled shear stress, necessitating an increase in the mean velocity gradient to maintain the balance of total shear stress. As a consequence, the positive LLM occurs at one grid away from the modification zone. Furthermore, along with the positive LLM, we can see an increased error of ${\varepsilon _{LD}}$ in the cases of WM2 and WM3.
	
	From another point of view, the LLM phenomenon arises from the incongruity between the resolved Reynolds shear stress and the modeled Reynolds shear stress. Figure \ref{fig:WMn-ratio} demonstrates the ratio of the resolved part to the total shear stress (the summation of the resolved and modeled parts) for the above four cases. The lower ratio within the modification zone compared with the DNS data indicates the real Reynolds number of the near-wall flow is lower than expected, which is supported by the lower velocity slope in the near-wall region observed in figure \ref{fig:WMn-models}(a).  The 'weak-turbulence region' can extend up to one layer beyond the modification zone. This causes the smaller SGS eddy viscosity calculated by the dynamic Smagorinsky model close to the modification zone and consequently leads to an upward shift of mean velocity profile at that position (known as the positive LLM phenomenon). In other words, for the WM2 case, the LLM phenomenon occurs at the third grid point away from the wall, while for the WM3 case, it occurs at the fourth grid point. 
	
	\subsection{Pressure gradient effect}\label{sec:Pressure}
	
	\begin{table}
		\captionsetup{justification=raggedright,singlelinecheck=true}
		\caption{\label{tab:WMpn-models}
			Comparison of wall shear stress among the wall model, the LES solver, and the DNS data at ${Re}_\tau=550$ with the WMpn ($n=0-3$) models.}
		\begin{ruledtabular}
			\begin{tabular}{lcccc}
				\textrm{Case}&
				\textrm{WMp0}&
				\textrm{WMp1}&
				\textrm{WMp2}&
				\textrm{WMp3} \\
				\colrule
				$\varepsilon_{LW}(\%)$ & -0.08 & -0.08 & 0.11 & 0.07 \\
				$\varepsilon_{LD}(\%)$ & -0.97 & -1.42 & -3.03 & -2.12 \\
			\end{tabular}
		\end{ruledtabular}
	\end{table}
	
	\begin{figure}
		\centering
		\begin{overpic}[width=0.48\linewidth
			]{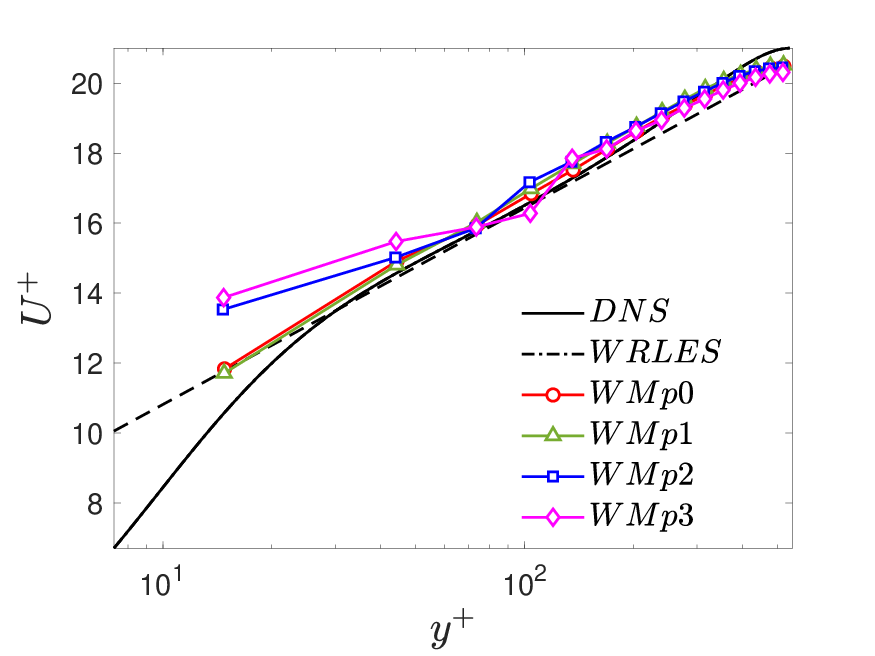}
			\put(7,73){($a$)}
			\put(30,60){\color{red}LLM region}
			\put(61,52){\color{red}\circle{20}}
		\end{overpic}
		\begin{overpic}[width=0.48\linewidth
			]{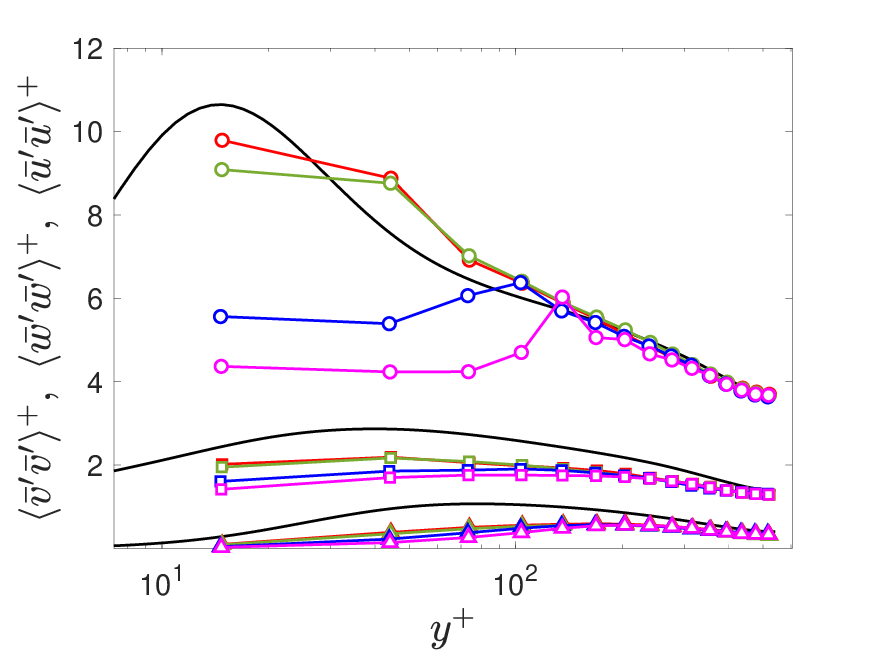}
			\put(7,73){($b$)}
		\end{overpic}
		\begin{overpic}[width=0.48\linewidth
			]{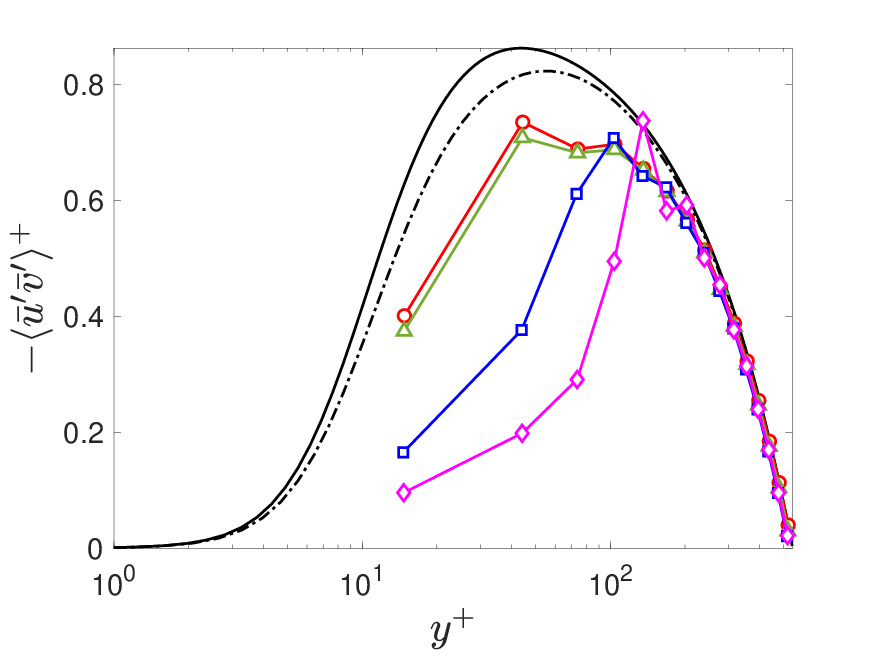}
			\put(7,73){($c$)}
		\end{overpic}
		\begin{overpic}[width=0.48\linewidth
			]{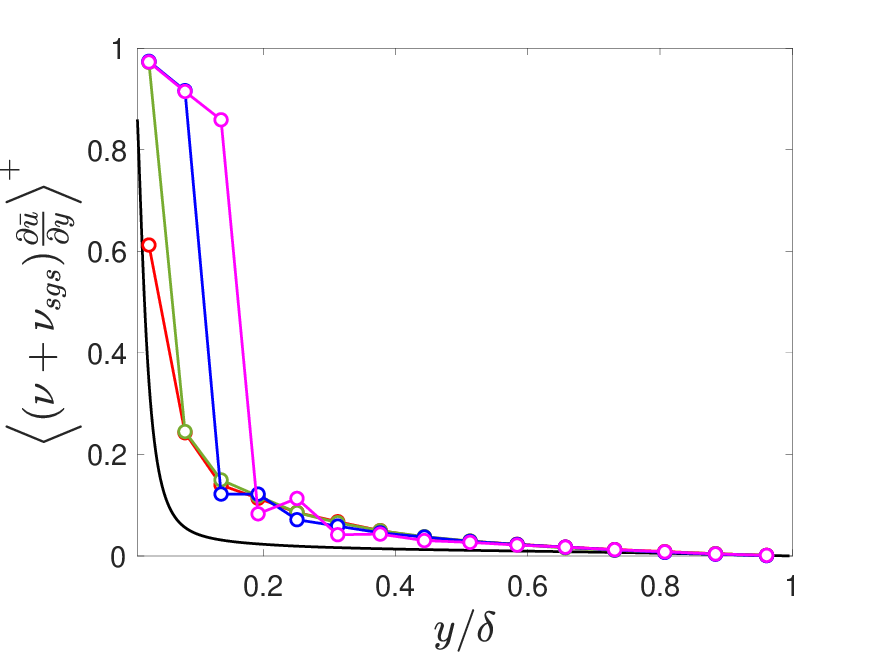}
			\put(7,73){($d$)}
		\end{overpic}
		\begin{overpic}[width=0.48\linewidth
			]{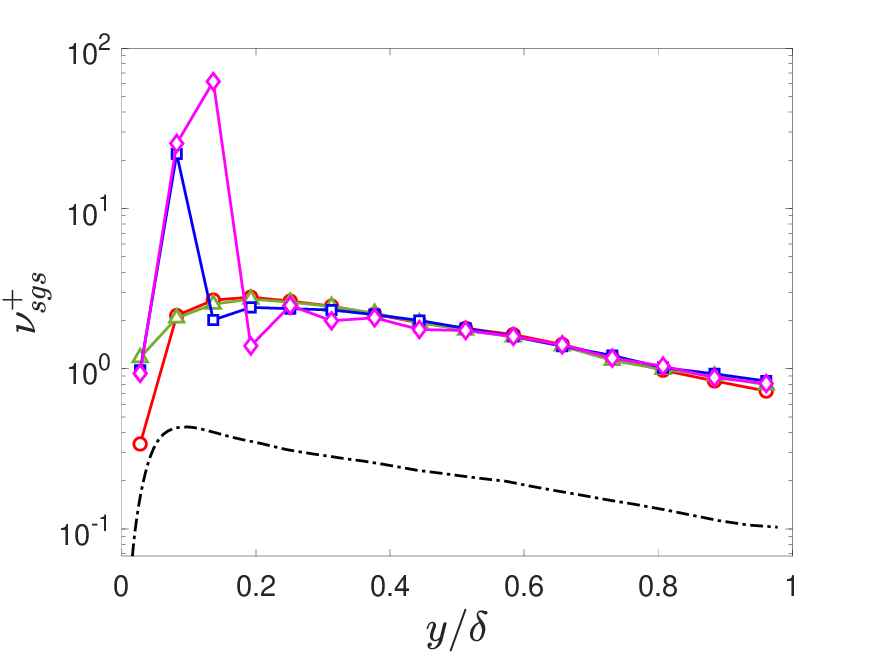}
			\put(7,73){($e$)}
		\end{overpic}
		\begin{overpic}[width=0.48\linewidth
			]{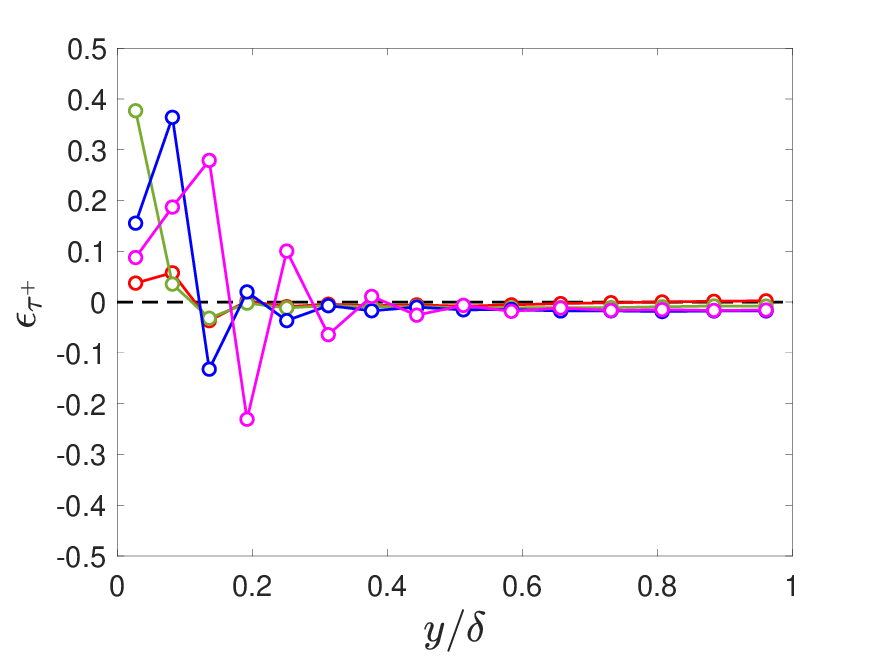}
			\put(7,73){($f$)}
		\end{overpic}\\
		\quad\\
		\captionsetup{justification=raggedright,singlelinecheck=true}
		\caption{(Colour online) Simulation results from the WMpn ($n=0-3$) models at ${Re}_\tau=550$ compared with DNS data from Lee and Moser \cite{lee_direct_2015} and the WRLES results. Refer to figure \ref{fig:WMn-models} for the detailed description of the figure caption.}
		\label{fig:WMpn-models}
	\end{figure}

	According to the discussion in $\S$\ref{sec:formation}, the oversimplification of the WMn model increases the risk of the positive LLM. In this subsection, the WMpn model is tested to study the pressure gradient effect on the positive LLM phenomenon.
	
	In table \ref{tab:WMpn-models}, it can be clearly seen that all $\left| {{\varepsilon _{LW}}} \right|$ are less than about 0.1$\%$, and all $\left| {{\varepsilon _{LD}}} \right|$ are less than about 3$\%$, which are better than the results of the WMn cases (compared with those in table \ref{tab:WMn-models}). Similar to figure \ref{fig:WMn-models}, we show the profiles of the mean velocity, the Reynolds normal stresses, the Reynolds shear stresses, the summation of the modeled Reynolds shear stress and the viscous stress, the SGS eddy viscosity and the deviation of the total shear stress in figure \ref{fig:WMpn-models}(a-f), respectively. By including the pressure gradient term, the overestimation of the modeled Reynolds shear stress and the total shear stress becomes less significant [see figure \ref{fig:WMn-models}(d,f) and figure \ref{fig:WMpn-models}(d,f)]. As a result, the near-wall fluctuations become more active, and the Reynolds normal stresses as well as the resolved Reynolds shear stress exhibit smoother distributions [see figure \ref{fig:WMn-models}(b,c) and figure \ref{fig:WMpn-models}(b,c)]. Consequently, the positive LLM phenomenon is alleviated in some extent, where the mean velocity profile in the WMp2 case (or the WMp3 case) is shifted less upward than that in the WM2 case (or the WM3 case).
	
	Although inclusion of the pressure gradient effect improves the prediction of the wall friction coefficient and low-order turbulence statistics, the issues present in the WMn model remain unresolved in the WMpn model. Because of the absence of the convection term, the total shear stress exceeds and thereby suppresses the fluctuations in the near-wall region. With the modification zone covering more layers of grids, this problem becomes more severe and causes the obvious positive LLM phenomenon in the WMp2 and WMp3 cases. In other words, the introduction of the pressure gradient term is not sufficient to ensure proportional coordination between the resolved and modeled Reynolds shear stresses.
	
	\subsection{Convection effect}\label{sec:convection}
	
	\begin{table}
		\captionsetup{justification=raggedright,singlelinecheck=true}
		\caption{\label{tab:WMpcn-models}
			Comparison of wall shear stress among the wall model, the LES solver, and the DNS data at ${Re}_\tau=550$ with the WMpcn ($n=0-3$) models.}
		\begin{ruledtabular}
			\begin{tabular}{lcccc}
				\textrm{Case}&
				\textrm{WMpc0}&
				\textrm{WMpc1}&
				\textrm{WMpc2}&
				\textrm{WMpc3} \\
				\colrule
				$\varepsilon_{LW}(\%)$ & -0.02 & 0.08 & 0.07 & -0.09 \\
				$\varepsilon_{LD}(\%)$ & -0.88 & -0.88 & 0.28 & 0.03 \\
			\end{tabular}
		\end{ruledtabular}
	\end{table}
	
	\begin{figure}
		\centering
		\begin{overpic}[width=0.48\linewidth
			]{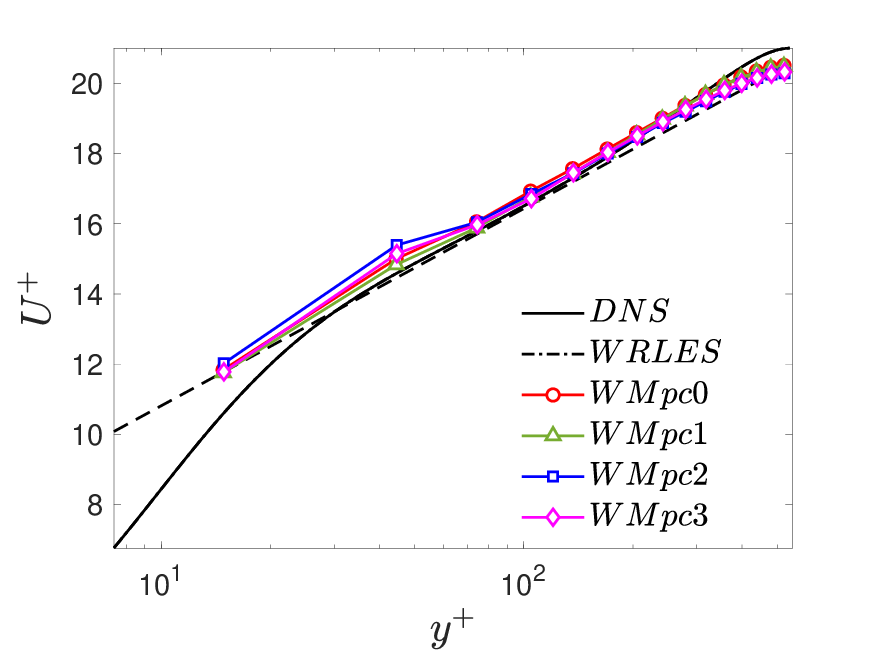}
			\put(7,73){($a$)}
		\end{overpic}
		\begin{overpic}[width=0.48\linewidth
			]{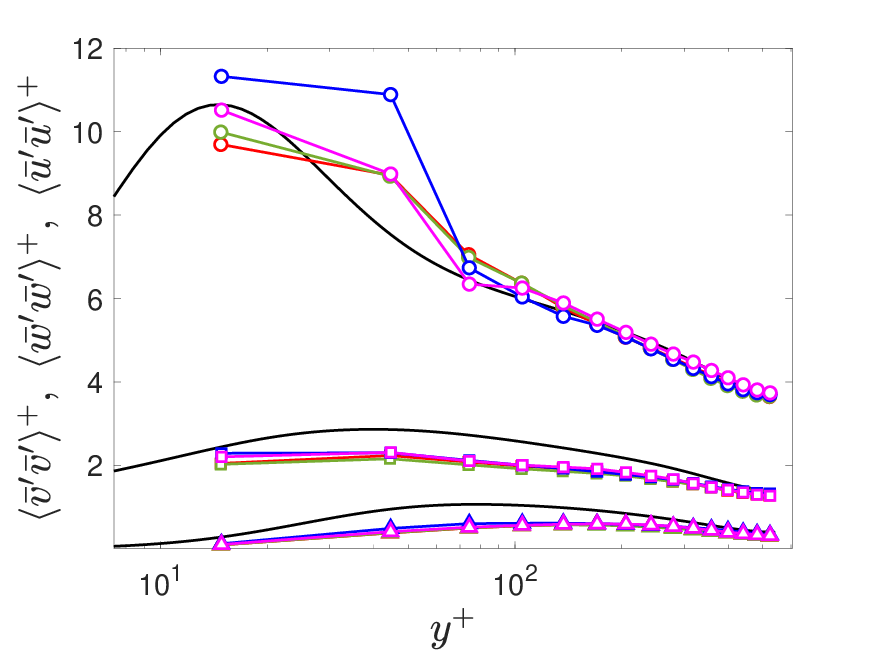}
			\put(7,73){($b$)}
		\end{overpic}
		\begin{overpic}[width=0.48\linewidth
			]{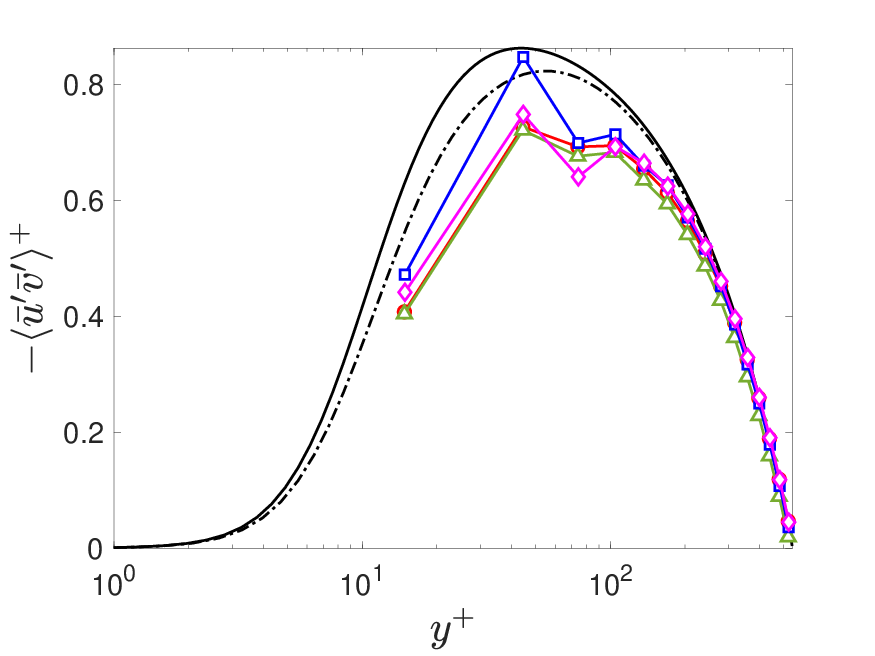}
			\put(7,73){($c$)}
		\end{overpic}
		\begin{overpic}[width=0.48\linewidth
			]{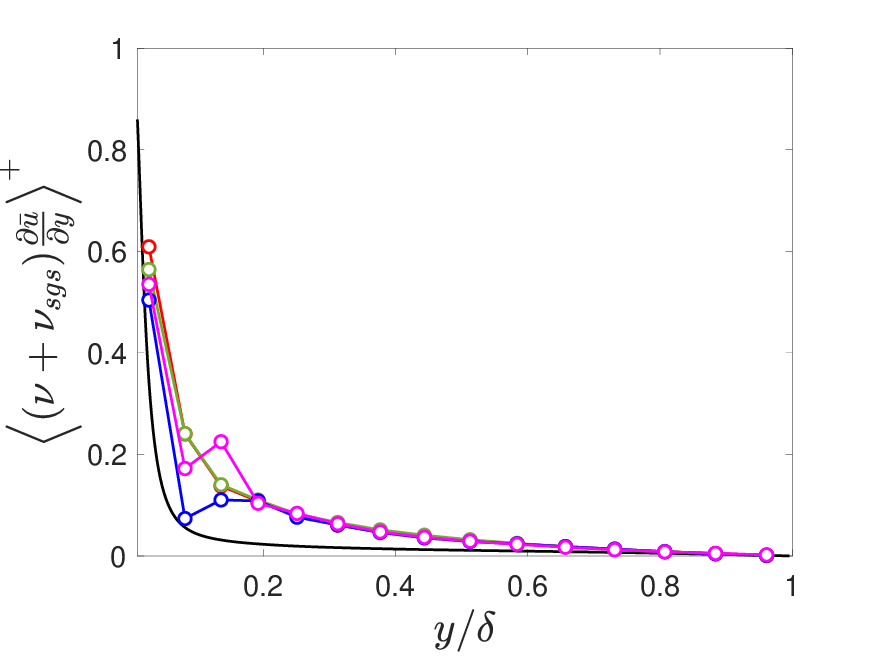}
			\put(7,73){($d$)}
		\end{overpic}
		\begin{overpic}[width=0.48\linewidth
			]{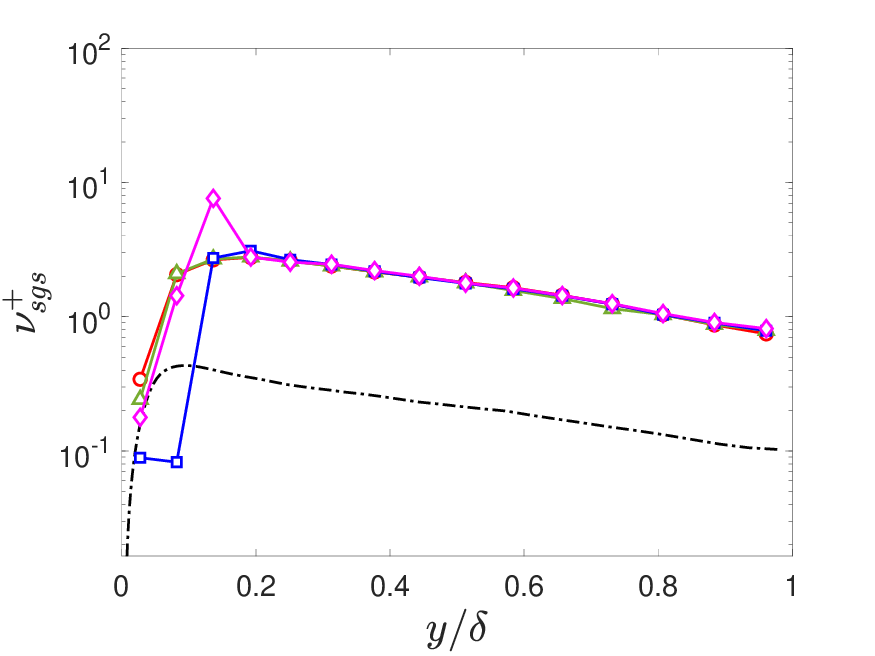}
			\put(7,73){($e$)}
		\end{overpic}
		\begin{overpic}[width=0.48\linewidth
			]{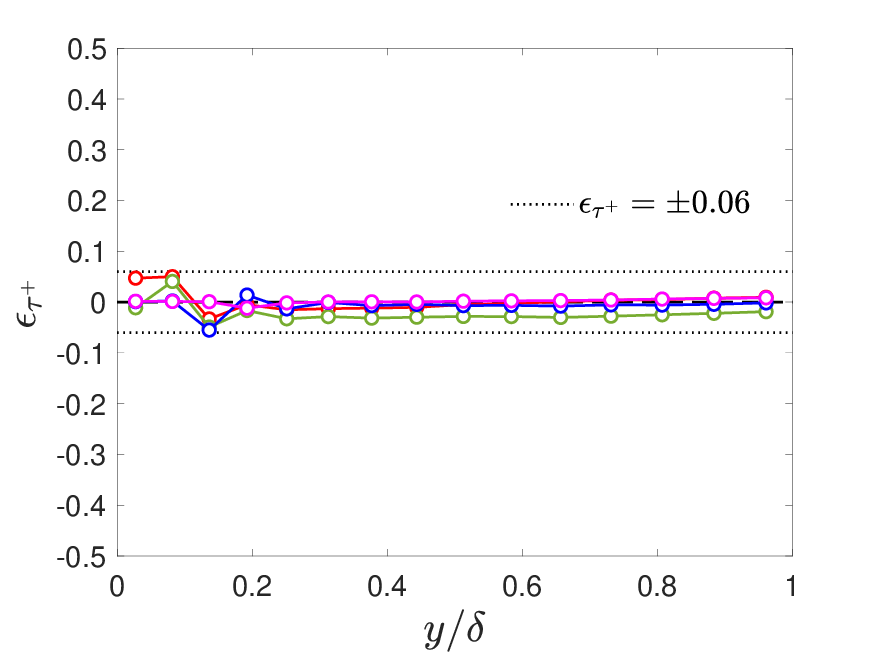}
			\put(7,73){($f$)}
		\end{overpic}\\
		\quad\\
		\captionsetup{justification=raggedright,singlelinecheck=true}
		\caption{(Colour online) Simulation results from the WMpcn ($n=0-3$) models at ${Re}_\tau=550$ compared with DNS data from Lee and Moser \cite{lee_direct_2015} and the WRLES results. Refer to figure \ref{fig:WMn-models} for the detailed description of the figure caption.}
		\label{fig:WMpcn-models}
	\end{figure}
	
	To further examine the convection effect, the WMpcn ($n=0-3$) model is then tested in this subsection, which is in essence the TSSC model as indicated in $\S$\ref{sec:wall}. Table \ref{tab:WMpcn-models} presents the comparison of wall shear stress calculated by the WMpcn model, the LES solver and the DNS data. It is satisfying to see that for each case, by introducing the convection effect, $\left| {{\varepsilon _{LW}}} \right|$ reduces to less than 0.1$\%$ and $\left| {{\varepsilon _{LD}}} \right|$ is less than 1$\%$. The SGS eddy viscosity calculated by the WMpcn models exhibits a more reasonable value in the modification zone, while the modeled Reynolds shear stress is predicted more accurately (see figure \ref{fig:WMpcn-models}d,e). Furthermore, as seen in figure \ref{fig:WMpcn-models}(f), the deviation of the total shear stress is dramatically reduced to less than 0.06 ${\tau _w}$, regardless of the number of layers in the modification zone. Therefore, the profiles of the resolved Reynolds stresses are distributed more reasonably, with no observation of unphysically suppressed profiles in the near-wall region. Additionally, the mean velocity profiles nearly coincide with each other and greatly match the expected logarithmic law. 
	
	For different numbers of layers in the modification zone, the errors of the wall shear stress change slightly, while the turbulence statistics collapse well in the outer layer. However, some discrepancies can be observed at the first few points away from the wall. In the near-wall region, the streamwise Reynolds normal stress is slightly higher for the WMpc2 model, and the profiles shown in figure \ref{fig:WMpcn-models}(c,d,e) are more jagged for the WMpc2 and WMpc3 models. This may be attributed to the fact that the fluctuation part of the SGS eddy viscosity is not considered by the WMpcn model. As a result, with the modification zone covering more layers, although the WMpcn model enforces more regions to satisfy the conversation of the total shear stress, it influences the flow structures in the near-wall region and makes the profiles of the resolved Reynolds shear stress more jagged. The effect of modification layers mentioned above is significant in the high-resolution grid ($\Delta {x^ + },\Delta {z^ + } \mathbin{\lower.3ex\hbox{$\buildrel<\over{\smash{\scriptstyle\sim}\vphantom{_x}}$}} 100$), but weak in the low-resolution grid ($\Delta {x^ + },\Delta {z^ + }\mathbin{\lower.3ex\hbox{$\buildrel>\over{\smash{\scriptstyle\sim}\vphantom{_x}}$}} 200$). More details of the effect of modification layers are presented in Appendix \ref{sec:appA}.
	
	\subsection{Revisiting the explanation and remedy of LLM}\label{sec:revisiting}
	In $\S$\ref{sec:formation}-\ref{sec:convection}, we exhibit the positive LLM caused by the exceeded total shear stress within the modification zone, and as a remedy, the convection effect must be introduced to ensure a reasonable distribution of the total shear stress in the near-wall region [compare figure\ref{fig:WMpcn-models}(f) with figure\ref{fig:WMn-models}(f) and figure\ref{fig:WMpn-models}(f)]. To elliminate the LLM phenomenon, it is essential to avoid an abrupt change of the velocity gradient, meaning that the ratio of the resolved and modeled Reynolds shear stresses should not vary too rapidly in the wall-normal direction. 
	
	Piomelli \textit{et al.} \cite{piomelli_innerouter_2003} used the Spalart-Allmaras (SA) model to resolve the flow in the inner layer, but found that the SA model tends to attenuate the Reynolds-stress containing structures and, therefore, makes the proportion of the resolved Reynolds shear stress less than expected. Accordingly, they added the stochastic force to generate rotational motions, which effectively supports a higher proportion of the resolved Reynolds shear stress and successfully alleviates the LLM phenomenon. However, they also pointed out that the backscatter model based on stochastic force has no physical justification, and the amplitude of stochastic force should be determined carefully and empirically.  On the contrary, the present model does not require empirical equations or free parameters.
	
	Shur \textit{et al.} \cite{shur_hybrid_2008} applied several semi-empirical corrections to the length scale ($d_{LES}$) in the DES model \citep{spalart_comments_1997} to make the SGS eddy viscosity have a steep decrease close to the wall. This results in a significant increase in the resolved part of the stress and a substantial suppression of the LLM. In this paper, by ensuring the conservation of the total shear stress by involving the convection effect, the modeled part of the stress can adaptively adjust itself according to the distribution of the resolved part, and thus the positive LLM can be effectively eliminated.
	
	Chen \textit{et al.} \cite{chen_reynolds-stress-constrained_2012} also realized the importance of the balance between the resolved and modeled parts of the Reynolds shear stress. They enforced a Reynolds-stress constraint (RSC) on the mean part of the modeled SGS stress to enforce the mean flow in the inner layer to satisfy the RANS solution, i.e., $\left\langle {\tau _{ij}^{LES}} \right\rangle  \equiv R_{ij}^{RANS} - R_{ij}^{LES}$, where the $R_{ij}^{RANS} = \left\langle {{u_i}{u_j}} \right\rangle  - \left\langle {{u_i}} \right\rangle \left\langle {{u_j}} \right\rangle $ and $R_{ij}^{LES} = \left\langle {{{\bar u}_i}{{\bar u}_j}} \right\rangle  - \left\langle {{{\bar u}_i}} \right\rangle \left\langle {{{\bar u}_j}} \right\rangle $ denote the total Reynolds shear stress from the RANS model and the resolved Reynolds shear stress from the LES solver, respectively. From this point of view, we can regard the present TSSC model as a total shear stress constraint on the modeled SGS stress. In fact, the TSSC model can also satisfy the equality $\left\langle {\tau _{ij}^{LES}} \right\rangle  \equiv R_{ij}^{RANS} - R_{ij}^{LES}$. Using the eddy-viscosity assumption for $\tau _{ij}^{LES}$, and decomposing $\bar u\bar v$ into the sum of $\left\langle {\bar u} \right\rangle \left\langle {\bar v} \right\rangle$ and $\bar{u}^{\prime}\bar{v}^{\prime}$, Eq.(\ref{eq:simplified-eq-x}) can be reorganized as:
	\begin{equation}
		\underbrace { - \left\langle {2{{\bar \nu }_{sgs}}{{\bar S}_{12}}} \right\rangle }_{\left\langle {\tau _{12}^{LES}} \right\rangle } = \underbrace { - \frac{{\left\langle {{\tau _w}} \right\rangle }}{\rho } - \frac{{\partial \left\langle {{{\bar p}^*}} \right\rangle }}{{\partial x}}y + \nu \frac{{\partial \left\langle {\bar u} \right\rangle }}{{\partial y}} - \left\langle {\bar u} \right\rangle \left\langle {\bar v} \right\rangle }_{R_{12}^{RANS}} - \underbrace {\left\langle {\bar u'\bar v'} \right\rangle }_{R_{12}^{LES}}.
	\end{equation}
	The equality $R_{12}^{RANS} =  - \frac{{\left\langle {{\tau _w}} \right\rangle }}{\rho } - \frac{{\partial \left\langle {{{\bar p}^*}} \right\rangle }}{{\partial x}}y + \nu \frac{{\partial \left\langle {\bar u} \right\rangle }}{{\partial y}} - \left\langle {\bar u} \right\rangle \left\langle {\bar v} \right\rangle $ can be proved as follows. Take the ensemble average of Eq.(\ref{eq:LES}) and obtain:
	\begin{equation}\label{eq:ensemble-LES}
		\frac{{\partial \left\langle {{{\bar u}_i}} \right\rangle }}{{\partial t}} + \frac{{\partial \left\langle {{{\bar u}_i}} \right\rangle \left\langle {{{\bar u}_j}} \right\rangle }}{{\partial {x_j}}} =  - \frac{{\partial \left\langle {{{\bar p}^*}} \right\rangle }}{{\partial {x_i}}} + \nu \frac{{{\partial ^2}\left\langle {{{\bar u}_i}} \right\rangle }}{{\partial {x_j}\partial {x_j}}} - \frac{{\partial R_{ij}^{RANS}}}{{\partial {x_j}}},
	\end{equation}
	where $R_{ij}^{RANS} = \left\langle {\bar \tau _{ij}^{LES,d}} \right\rangle  + \left\langle {{{\bar u}_i}^\prime {{\bar u}_j}^\prime } \right\rangle $. For the fully developed turbulent channel flow, it is reasonable to omit the unsteady term, streamwise gradient term, and spanwise gradient term. By integrating the streamwise momentum equation of Eq.(\ref{eq:ensemble-LES}) from $0$ to $y$ along the wall-normal direction, we can obtain $R_{12}^{RANS} =  - \frac{{\left\langle {{\tau_w}} \right\rangle }}{\rho } - \frac{{\partial \left\langle {{{\bar p}^*}} \right\rangle }}{{\partial x}}y + \nu \frac{{\partial \left\langle {\bar u} \right\rangle }}{{\partial y}} - \left\langle {\bar u} \right\rangle \left\langle {\bar v} \right\rangle $. Moreover, the present TSSC model does not additionally model the fluctuation part of the SGS eddy viscosity, which is different from the RSC-LES model of Chen \textit{et al.} \cite{chen_reynolds-stress-constrained_2012}. Regarding the requirement of grid resolution, there is no limitation for the wall-normal grid resolution in the TSSC model (see $\S$\ref{sec:effects-grid}), while the wall-normal grid resolution is better to satisfy $y_1^+\lesssim 1$ for the RSC-LES model Chen \textit{et al.} \citep{chen_reynolds-stress-constrained_2012}.
	
	Kalitzin \textit{et al.} \cite{kalitzin_anear-wall_2007} derived a RANS-like eddy-viscosity corrected with the resolved Reynolds shear stress. Incorporation of the resolved Reynolds shear stress improves the distribution of the calculated SGS eddy viscosity in the near-wall region and satisfies the RSC constraint, successfully removing the LLM. In their model, the RANS viscosity and the mean velocity gradient are obtained from the look-up table, which imposes a certain limitation on its practicability.
	
	In summary, to explain the positive or negative LLM, we suppose that it is the incongruity of the resolved and modeled stresses that induces an abrupt increase or decrease of the velocity gradient at the interface between the wall-model region and the outer region. The present TSSC model can promise the rationality of the ratio of the resolved part to the modeled part of the Reynolds shear stress in the near-wall region. As a result, there is no need to explicitly add stochastic force or adjust the length scale to promote the generation of the Reynolds-stress containing structures. In other words, the present model does not require empirical functions or free parameters. The proposed TSSC model can also guarantee the mean flow in the near-wall region to satisfy the RANS solution, as the RSC-LES model does. In addition, it has good performance with relatively coarse grids ($\Delta {x^ + },\Delta {z^ + } \lesssim 500$, and the first grid height can be more than 30 wall units, as seen in $\S$\ref{sec:effects-grid}), which is computationally tractable for high-Reynolds number wall turbulence.
	
	\section{Implementation of the TSSC model at high Reynolds numbers}\label{sec:TSSC}

	\begin{table}
		\captionsetup{justification=raggedright,singlelinecheck=true}
		\caption{\label{tab:test-TSSC}
			Comparison of the wall shear stress from the DNS data, LES solver, and TSSC model under different coverages of the modification zone ($n=0-3$) at different Reynolds numbers (${Re}_{\tau}=1000,\ 2000,\ 4200$) in the fully developed turbulent channel flow. The DNS data of ${Re}_{\tau}=1000,2000$ is from Lee and Moser \cite{lee_direct_2015}, and the DNS data of ${Re}_\tau=4200$ is from Lozano-Dur\'an and Jim\'enez\cite{lozano-duran_effect_2014}. The grids are uniform in horizontal directions and stretched in the wall-normal direction.}
		\begin{ruledtabular}
			\begin{tabular}{cccccccc}
				\textrm{Case}&
				\textrm{$Re_{\tau}$}&
				\textrm{$N_x\times N_y\times N_z$}&
				\textrm{$\mathrm{\Delta}_x^+$}&
				\textrm{$\mathrm{\Delta}_z^+$}&
				\textrm{$\mathrm{\Delta}_{y,min}^+,\ \mathrm{\Delta}_{y,max}^+$}&
				\textrm{$\varepsilon_{LW}(\%)$}&
				\textrm{$\varepsilon_{LD}(\%)$} \\
				\colrule
				WMpc0 & 1000 & $32\times30\times32$ & 196 & 98 & [55,78] & -0.10 & 0.87 \\
				WMpc1 & 1000 & $32\times30\times32$ & 196 & 98 & [55,78] & -0.34 & 0.76 \\
				WMpc2 & 1000 & $32\times30\times32$ & 196 & 98 & [55,78] &  0.16 & 2.02 \\
				WMpc3 & 1000 & $32\times30\times32$ & 196 & 98 & [55,78] &  0.01 & 1.06 \\
				WMpc0 & 2000 & $32\times30\times32$ & 393 & 196 & [109,156] & -0.13 & 2.87 \\
				WMpc1 & 2000 & $32\times30\times32$ & 393 & 196 & [109,156] & -0.33 & 3.41 \\
				WMpc2 & 2000 & $32\times30\times32$ & 393 & 196 & [109,156] &  0.16 & 3.57 \\
				WMpc3 & 2000 & $32\times30\times32$ & 393 & 196 & [109,156] & -0.08 & 2.30 \\
				WMpc0 & 4200 & $64\times40\times32$ & 412 & 412 & [137,274] & -0.24 & 3.73 \\
				WMpc1 & 4200 & $64\times40\times32$ & 412 & 412 & [137,274] &  0.10 & 3.97 \\
				WMpc2 & 4200 & $64\times40\times32$ & 412 & 412 & [137,274] &  0.05 & 3.98 \\
				WMpc3 & 4200 & $64\times40\times32$ & 412 & 412 & [137,274] &  0.06 & 3.67 \\
			\end{tabular}
		\end{ruledtabular}
	\end{table}
	
	\begin{figure}
		\centering
		\begin{overpic}[width=0.49\linewidth
			]{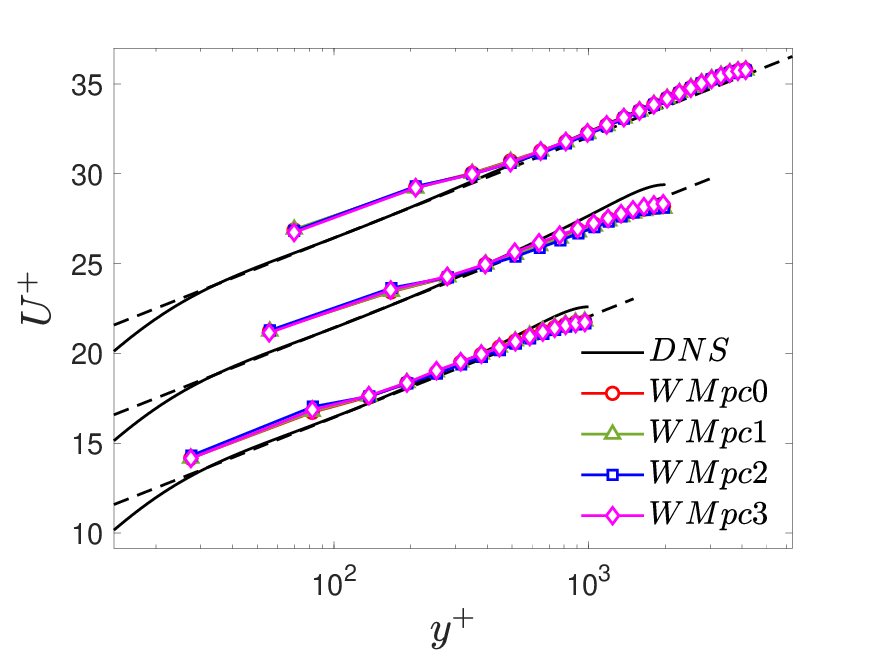}
			\put(7,73){($a$)}
		\end{overpic}
		\begin{overpic}[width=0.49\linewidth
			]{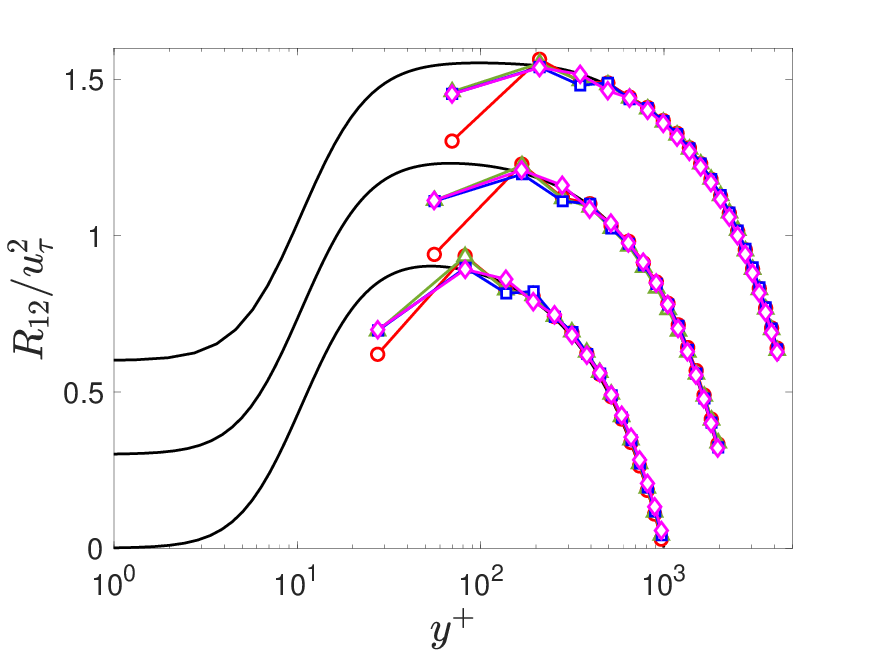}
			\put(7,73){($b$)}
		\end{overpic}\\
		\quad\\
		\captionsetup{justification=raggedright,singlelinecheck=true}
		\caption{(Colour online) Comparison of the simulation results of the cases listed in table~\ref{tab:test-TSSC} and the reference DNS data from \cite{lee_direct_2015} for ${Re}_\tau=1000,2000$ and \cite{lozano-duran_effect_2014} for ${Re}_\tau=4200$. (a) The mean velocity profile $U=\langle \bar{u} \rangle$, where the black dashed line denotes the logarithmic law $U^+=\frac{1}{0.41}ln(y^+)+5.2$. The lines are shifted upward by 5 and 10 units for ${Re}_\tau=2000$ and 4200, respectively. (b) The total Reynolds shear stress, which are shifted upward by 0.3 and 0.6 units for ${Re}_\tau=2000$ and 4200, respectively.}
		\label{fig:test-TSSC}
	\end{figure}
	
	In this section, we apply the TSSC model in turbulent channel flows at different Reynolds numbers (${Re}_\tau=1000, 2000, 4200$), and for each Reynolds number, we test different coverages of the modification zone ($n=0-3$). The wall shear stress error and grid setting of each case are listed in table~\ref{tab:test-TSSC}. The corresponding mean velocity profile and total Reynolds shear stress ($R_{12}=\langle -\bar{u}^\prime \bar{v}^\prime + \nu_{sgs}\frac{\partial\bar{u}}{\partial{y}} \rangle$) are displayed in figure~\ref{fig:test-TSSC}. As expected, all $\left|\varepsilon_{LW}\right|$ are less than 0.5$\%$, indicating a good feedback of wall shear stress from the wall model to the LES solver in the mean sense. Additionally, for each case, the total shear stress profile aligns with the expected one ($\frac{{{\tau _{total}}}}{{{\tau _{w,WMLES}}}} = 1 - \frac{y}{\delta }$) with a deviation of less than 0.06 ${\tau _{w,WMLES}}$, except for the first grid in the WMpc0 model (not shown here). This occurs because, for the WMpc0 model, the conservation of the total shear stress is not satisfied even at the first grid, and the Reynolds shear stress also deviates from the expected value (see figure~\ref{fig:test-TSSC}b). For each case in table~\ref{tab:test-TSSC}, $\left|\varepsilon_{LD}\right|$ is less than 4$\%$. As seen in figure~\ref{fig:test-TSSC}(a,b), the positive LLM is completely eliminated and, what is more, the total Reynolds shear stress is consistent with the DNS data except for the first grid. The above results confirm the accuracy and robustness of the TSSC model, irrespective of the Reynolds number and the number of layers in the modification zone. Furthermore, it is also important to test the effects of grid aspect ratio and grid resolution on the wall model, especially at high Reynolds numbers, which are discussed below.
	
	\subsection{Effects of the grid aspect ratio and grid resolution}\label{sec:effects-grid}
	
	For the TSSC model, it is found that when the grid remains unchanged, the deviation of the predicted wall shear stress to the actual value is increased with the increasing Reynolds number (not shown here). In other words, to obtain a reasonable wall shear stress, the grid resolution may not obey the outer-layer scaling as suggested by Larsson \textit{et al.} \cite{larsson_large_2016} and Chapman \cite{chapman_computational_1979}. This is because in the present wall model, the nonlinear convection term is calculated from the LES solver at the last time step. If the grid resolution is too low (in the inner-layer scale), the predicted convection term will deviate significantly from the actual value and, therefore, affect the performance of the wall model. According to our tests, not only the grid resolution, but also the grid aspect ratio in the wall-model region have influences on the results of the wall model. In the following, these two factors will be discussed separately.
	
	
	\begin{table}
		\captionsetup{justification=raggedright,singlelinecheck=true}
		\caption{\label{tab:grid-aspect}
			Comparison of wall shear stress errors with different grid resolutions from the WMpc3 model applied in the turbulent channel flow at ${Re}_\tau=4200$. The normal grid spacing $\left(\mathrm{\Delta}_{y,mat}^+\right)$ is uniform under the matching layer, and subsequently, it is stretched up to the channel center. For the cases of grid1a to grid8a, $\Delta _x^ +  = \Delta _z^ + $ and $\Delta_x^+/\Delta_{y,mat}^+$ ranges from 1 to 8. For the gird4b case, the number of spanwise grids is twice that of the grid4a case.}
		\begin{ruledtabular}
			\begin{tabular}{ccccc}
				\textrm{Case}&
				\textrm{$N_x\times N_y\times N_z$}&
				\textrm{$\mathrm{\Delta}_x^+:\mathrm{\Delta}_{y,mat}^+:\mathrm{\Delta}_z^+$}&
				\textrm{$\varepsilon_{LW}(\%)$}&
				\textrm{$\varepsilon_{LD}(\%)$} \\
				\colrule
				grid1a & $64\times 20\times 32$ & 1:1:1 & 0.13 & 2.08 \\
				grid2a & $64\times 30\times 32$ & 2:1:2 & 0.04 & 2.46 \\
				grid3a & $64\times 40\times 32$ & 3:1:3 & 0.06 & 3.67 \\
				grid4a & $64\times 50\times 32$ & 4:1:4 & -0.11 & 4.86 \\
				grid4b & $64\times 50\times 64$ & 4:1:2 & -0.08 & 6.74 \\
				grid5a & $64\times 60\times 32$ & 5:1:5 & 0.14 & 5.31 \\
				grid6a & $64\times 70\times 32$ & 6:1:6 & 0.10 & 5.34 \\
				grid7a & $64\times 80\times 32$ & 7:1:7 & -0.10 & 4.80 \\
				grid8a & $64\times 90\times 32$ & 8:1:8 & -0.05 & 3.64 \\
			\end{tabular}
		\end{ruledtabular}
	\end{table}
	
	\begin{figure}
		\centering
		\begin{overpic}[width=0.49\linewidth
			]{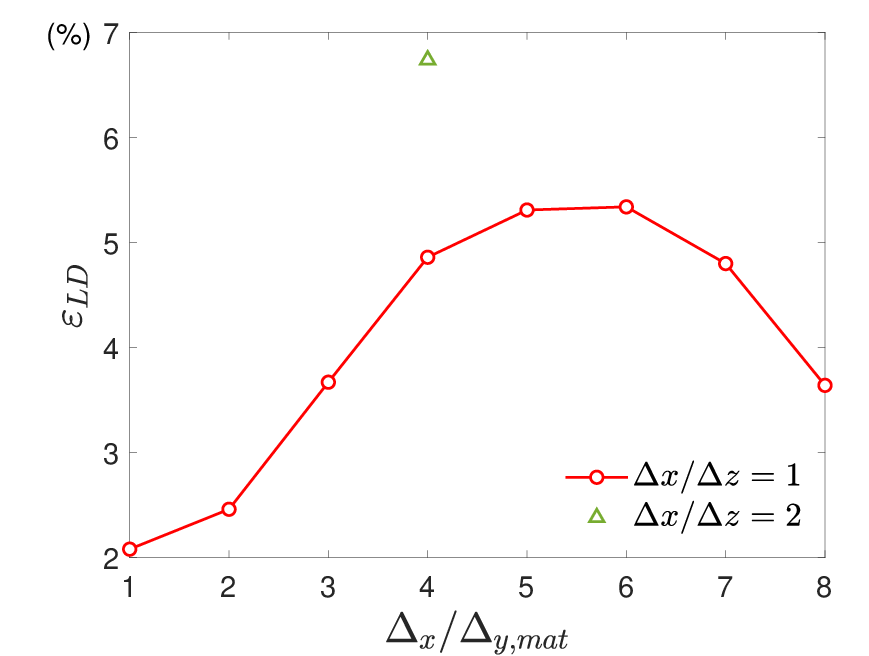}
			\put(18,63){($a$)}
		\end{overpic}
		\begin{overpic}[width=0.49\linewidth
			]{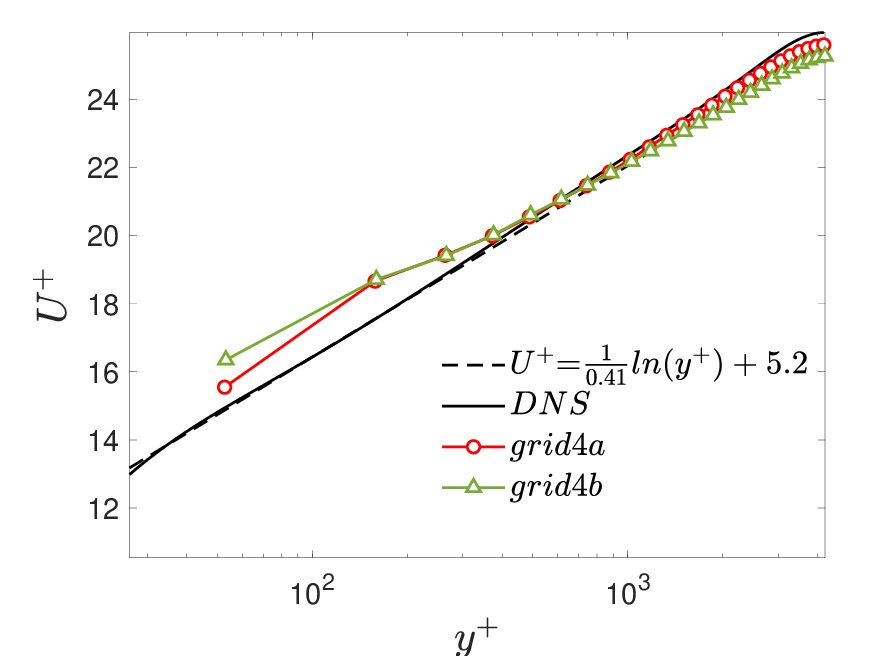}
			\put(18,63){($b$)}
		\end{overpic}
		\\
		\quad\\
		
		\captionsetup{justification=raggedright,singlelinecheck=true}
		\caption{Simulation results of different grids at ${Re}_\tau=4200$ for the WMpc3 model: (a) comparison of the wall shear stress errors with different ratios of the streamwise grid spacing to the normal grid spacing (below the matching layer); (b) comparison of the mean streamwise velocity profiles between the cases of grid4a and grid4b. Refer to the table~\ref{tab:grid-aspect} for the details of the grid settings.}
		\label{fig:gridAspect-12}
	\end{figure}
	
	\begin{figure}
		\centering
		\begin{overpic}[width=0.47\linewidth
			]{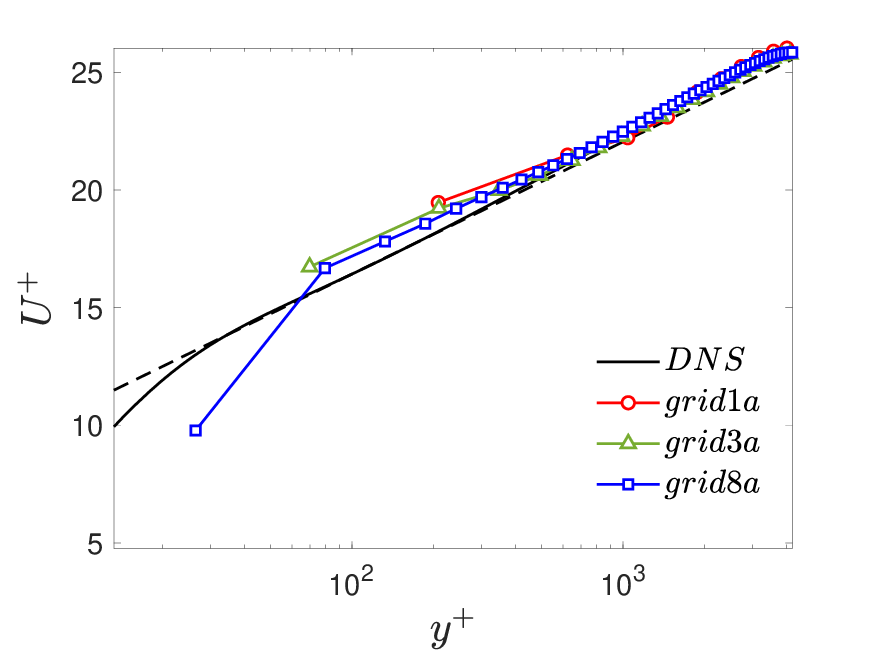}
			\put(7,73){($a$)}
		\end{overpic}
		\begin{overpic}[width=0.47\linewidth
			]{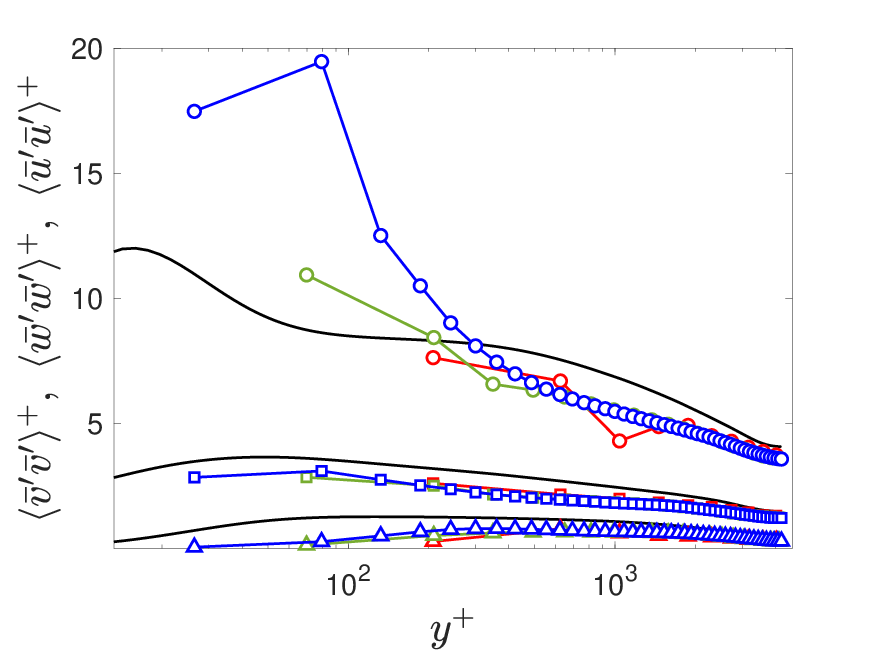}
			\put(7,73){($b$)}
		\end{overpic}\\
		\begin{overpic}[width=0.47\linewidth
			]{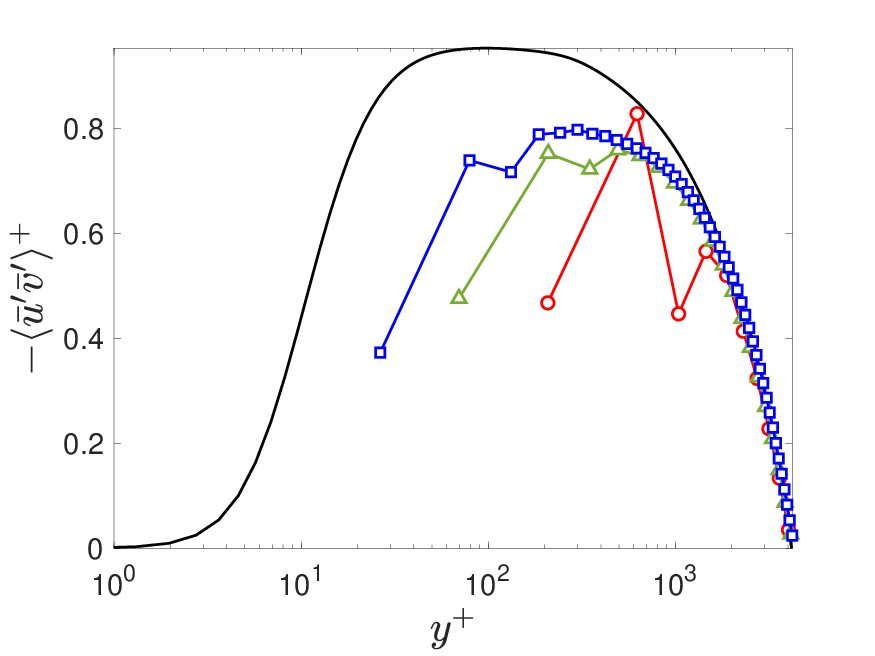}
			\put(7,73){($c$)}
		\end{overpic}
		\begin{overpic}[width=0.47\linewidth
			]{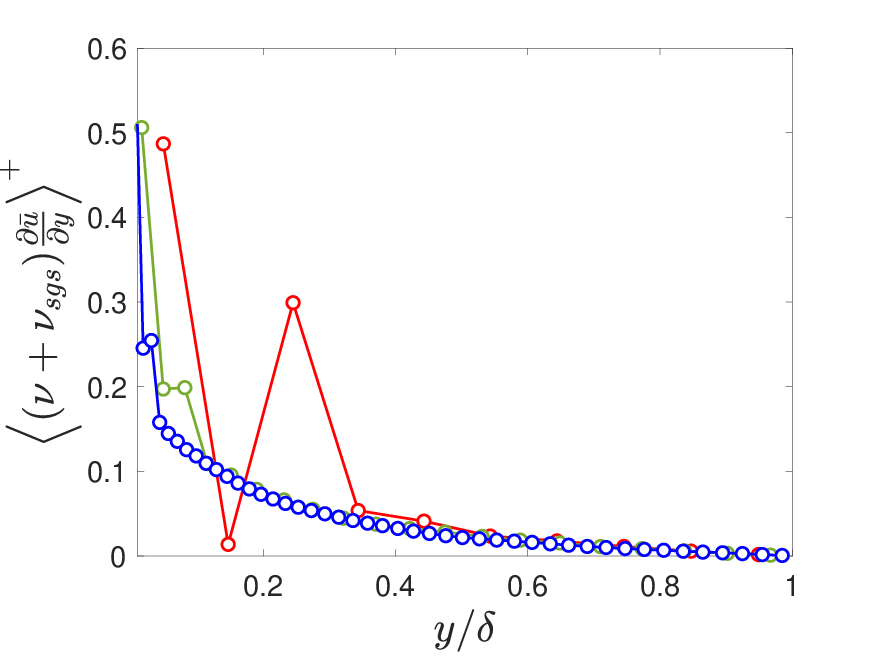}
			\put(7,73){($d$)}
		\end{overpic}
		\begin{overpic}[width=0.47\linewidth
			]{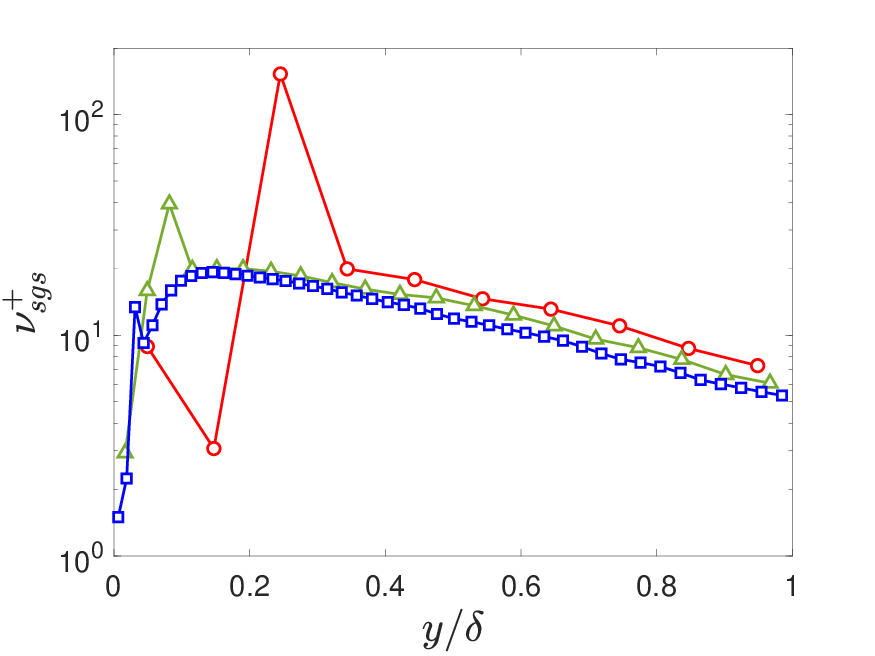}
			\put(7,73){($e$)}
		\end{overpic}
		\begin{overpic}[width=0.47\linewidth
			]{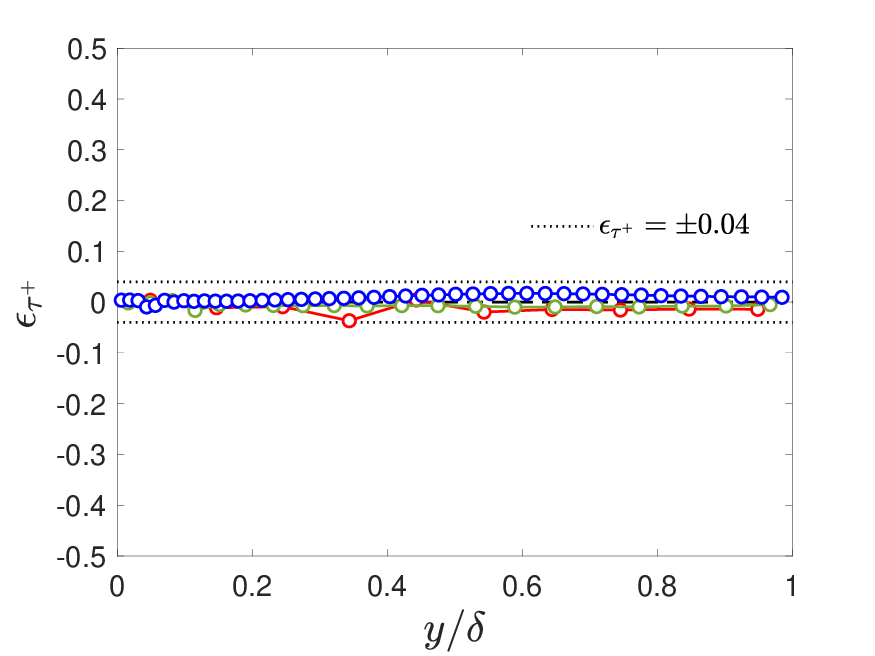}
			\put(7,73){($f$)}
		\end{overpic}\\
		\quad\\
		\captionsetup{justification=raggedright,singlelinecheck=true}
		\caption{Simulation results of three different grids at ${Re}_\tau=4200$ compared with DNS data from \cite{lozano-duran_effect_2014}. Refer to table~\ref{tab:grid-aspect} for the details of grid settings of the grid1a, grid3a and grid8a cases, and refer to figure \ref{fig:WMn-models} for the detailed description of the figure caption.}
		\label{fig:gridAspect-abcd}
	\end{figure}
	
	Table~\ref{tab:grid-aspect} shows the cases with several different grid settings for the WMpc3 model at $Re_\tau=4200$. The corresponding errors of the predicted wall shear stress obtained from the LES solver are plotted in figure~\ref{fig:gridAspect-12}(a). In previous studies of turbulent channel flow using the DNS method, the streamwise-to-spanwise grid aspect ratio was typically chosen between 1.4 and 2.5 \citep{lee_direct_2015,lozano-duran_effect_2014,hoyas_scaling_2006,bernardini_velocity_2014}. Here we consider two grid aspect ratios, i.e., $\Delta_{x}/\Delta_{z}=1$ and 2, as in the cases of grid4a and grid4b, respectively. It is seen in table~\ref{tab:grid-aspect} that the case of grid4a has a smaller ${\varepsilon _{LD}}$, which can be attributed to the fact that the mean velocity profile above the logarithmic layer is much closer to the DNS data. As a result, the streamwise-to-spanwise grid aspect ratio mainly affect the accuracy of LES solver to resolve the flow field in the outer region but not the performance of the TSSC model. For simplicity, we remain $\Delta_{x}/\Delta_{z}=1$ and change $\Delta_{x}/\Delta_{y,mat}$ from 1 to 8 to examine the effect of the streamwise-to-normal grid aspect ratio, where $\Delta_{y,mat}$ represents the uniform normal grid spacing below the matching layer. From figure~\ref{fig:gridAspect-12}(a), it is observed that the variation of the wall shear stress error with the streamwise-to-normal grid aspect ratio is not monotonic, consistent with the LES results of Nishizawa \textit{et al.} \cite{nishizawa_influence_2015}. Since the grid1a case has the lowest computational cost and the smallest error ${\varepsilon _{LD}}$, $\Delta_{x}/\Delta_{y,mat}=1$ is the best choice. Furthermore, to check whether the near-wall flow field is reasonably resolved, three cases ($\Delta_{x}/\Delta_{y,mat}=1,3,8$) are selected to compare the low-order turbulence statistics, as shown in figure~\ref{fig:gridAspect-abcd}. For the grid1a case, obvious oscillations are observed in the first few grid points away from the wall, especially for the resolved Reynolds shear stress and the modeled Reynolds shear stress, indicating a poor prediction of the flow structures in the near-wall region. With the increasing aspect ratio $\Delta_{x}/\Delta_{y,mat}$ (see the cases of grid3a and grid8a), the oscillations are alleviated, indicating a smoother transition for the flow structures from the wall-model region to the outer layer. When the mesh shape is excessively flat, that is, $\Delta_{x}/\Delta_{y,mat}=8$, the velocity at the first grid is significantly lower, and the resolved streamwise Reynolds normal stress at the first few grids is significantly higher than the DNS data. Accordingly, we suggest $2\leq\Delta_{x}/\Delta_{y,mat}\leq4$ based on a trade-off between the computational cost and the prediction accuracy of the wall shear stress and the near-wall flow fields.
	
	
	\begin{table}
		\captionsetup{justification=raggedright,singlelinecheck=true}
		\caption{\label{tab:grid-resolution}
			Comparison of wall shear stress errors with different Reynolds numbers and different grid resolutions for the WMpc3 model at ${Re}_\tau=4200$. The normal grid spacing is uniform under the matching layer ($\Delta _{y,mat}^ +$), and subsequently stretched up to the channel center. All the cases listed here satisfy $\Delta {x^ + } = \Delta {z^ + }$ and $\Delta_x^+/\Delta_{y,mat}^+= 3$.}
		\begin{ruledtabular}
			\begin{tabular}{ccccc}
				\textrm{$Re_\tau$}&
				\textrm{$N_x\times N_y\times N_z$}&
				\textrm{$\mathrm{\Delta}_x^+,\mathrm{\Delta}_y^+,\mathrm{\Delta}_z^+$}&
				\textrm{$\varepsilon_{LW}(\%)$}&
				\textrm{$\varepsilon_{LD}(\%)$} \\
				\colrule
				4200 & $32\times 30\times 16$ (G1) & 825,275-284,825 & 0.18 & 5.62 \\
				4200 & $64\times 40\times 32$ (G2) & 412,137-274,412 & 0.06 & 3.67 \\
				4200 & $128\times 60\times 64$ (G3) & 206,69-205,206 & -0.11 & -0.05 \\
				8000 & $128\times 60\times 64$ & 393,131-391,393 & -0.12 & 5.86 \\
				16000 & $256\times 80\times 128$ & 393,131-681,393 & 0.29 & 7.83 \\
				20000 & $256\times 80\times 128$ & 490,163-853,490 & -0.43 & 8.80 \\
			\end{tabular}
		\end{ruledtabular}
	\end{table}
	
	\begin{figure}
		\centering
		\begin{overpic}[width=0.48\linewidth
			]{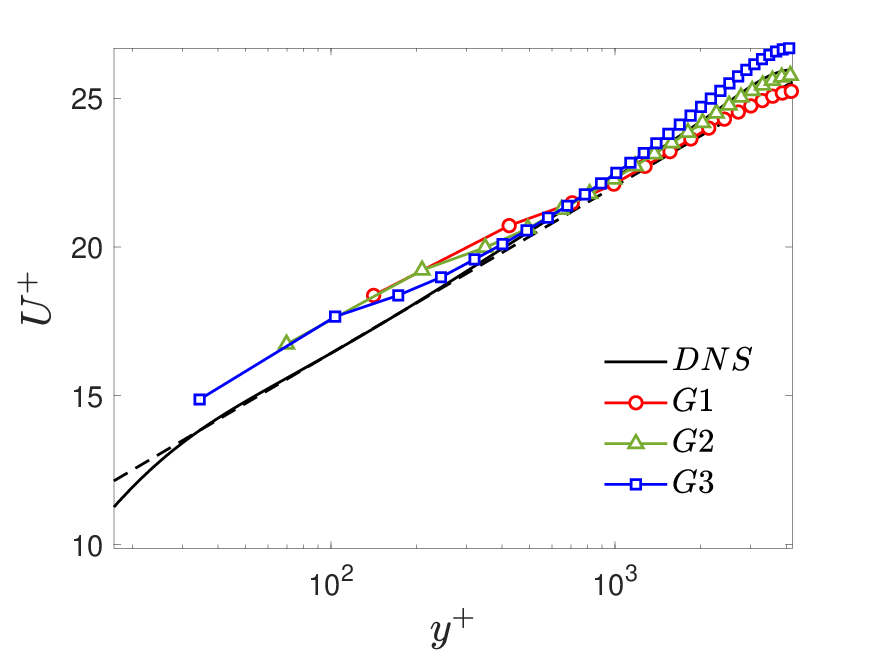}
			\put(7,73){($a$)}
		\end{overpic}
		\begin{overpic}[width=0.48\linewidth
			]{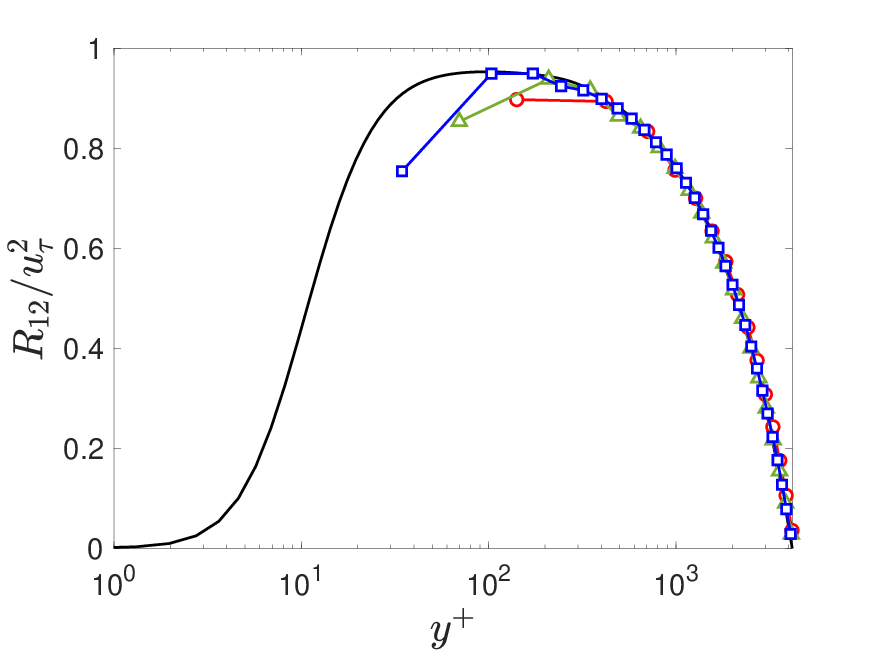}
			\put(7,73){($b$)}
		\end{overpic}\\
		\quad\\
		\captionsetup{justification=raggedright,singlelinecheck=true}
		\caption{Simulation results of three different grids at ${Re}_\tau=4200$ compared with DNS data of \cite{lozano-duran_effect_2014}: (a) mean velocity profiles; (b) total Reynolds shear stress. Refer to table~\ref{tab:grid-resolution} for the details of grid settings of the G1, G2 and G3 cases.}
		\label{fig:gridResolution}
	\end{figure}
	
	Furthermore, we fix $\Delta_x^+ / \Delta_{z}^+=1$, $\Delta_x^+/\Delta_{y,mat}^+=3$, and discuss the streamwise/spanwise grid resolution requirement for the TSSC model. As listed in table~\ref{tab:grid-resolution}, we choose three grid resolutions ($\Delta_x^+=\Delta_{z}^+\approx 800,400,200$; to be named G1, G2 and G3, respectively) for simulation of the turbulent channel flow at $Re_\tau=4200$. For the mean velocity profile (figure~\ref{fig:gridResolution}a), all the three cases are very close to the DNS data, although the slope of the profile above the logarithmic layer has slight discrepancy for the G1 and G3 cases. For the total Reynolds shear stress (figure~\ref{fig:gridResolution}b), all the three cases collapse well with the DNS data except for the first grid. As a conclusion, the low-order statistics can be predicted well (in the inner-layer scale) with relatively low streamwise/spanwise grid resolutions ($\Delta_x^+ = \Delta_{z}^+\lesssim 800$). However, as the grid spacing increases, the wall shear stress error ${\varepsilon _{LD}}$ actually increases (see table~\ref{tab:grid-resolution}). Therefore, to ensure that the error ${\varepsilon _{LD}}$ is less than 10$\%$ for a relatively wide range of Reynolds number, we use the WMpc3 model to simulate flows up to $Re_\tau=20000$ and find that $\Delta_x^+ = \Delta_{z}^+\lesssim 500$ meets the grid resolution requirement.
	
	In summary, it is suggested that the grid aspect ratio and grid resolution meet the criteria of $2\leq\Delta_{x}/\Delta_{y,mat}\leq 4$ and $\Delta_x^+ = \Delta_{z}^+\lesssim 500$, which can ensure that the wall shear stress error ($\varepsilon_{LD}$) remains less than 10$\%$ and allow a reasonable prediction of the near-wall flow field when simulating turbulent channel flow up to at least $Re_\tau=20000$.
	
	\section{Conclusions}\label{sec:conclusions}
	In the present study, unlike the canonical wall shear stress model, we explore the possibility of using the calculated wall shear stress to correct the SGS eddy viscosity in the near-wall region. By comparing three different SGS-EVM models, we found that the positive LLM phenomenon arises when the total shear stress conversation (${\tau ^ + } = {\left\langle {\left( {\nu  + {{\bar \nu }_{sgs}}} \right)\frac{{\partial \bar u}}{{\partial y}}} \right\rangle ^ + } - {\left\langle {\bar u\;\bar v} \right\rangle ^ + } = \left( {1 - \frac{y}{\delta }} \right)$) is not satisfied in the modification zone. To ensure the conservation of the total shear stress, the convection effect should be considered while the effect of pressure gradient is comparatively weaker. Although the simulation results do not show significant improvement with an increased modification zone, this approach provides potential improvements for the wall shear stress model. Our work points out the importance of the TSSC constraint when correcting the SGS eddy viscosity in the near-wall region. The proposed TSSC model was tested in turbulent channel flows at Reynold number up to ${{\mathop{\rm Re}\nolimits} _\tau } = 4200$. The present model shows good performance: the deviation of the wall shear stress from the DNS data is less than 4$\%$; the positive LLM is effectively eliminated and the low-order turbulence statistics agree well with the DNS and WRLES data. Notably, the TSSC model is derived directly from the streamwise momentum equation, without dependence on empirical functions or free parameters, enhancing its user-friendliness for engineers and suggesting its potential robustness for simulating a broader range of complex flows. We also discussed the impact of grid aspect ratio and the requirement of grid resolution. It is recommended that the grid aspect ratio and grid resolution should meet the criteria of $\Delta _x^ +  = \Delta _z^ +  \mathbin{\lower.3ex\hbox{$\buildrel<\over{\smash{\scriptstyle\sim}\vphantom{_x}}$}} 500$ and $2\leq\Delta_x/\Delta_{y,mat}\leq 4$ for the present TSSC model, ensuring both the acceptable wall shear stress error (${\varepsilon _{LD}}$) and the reasonable prediction of the near-wall flow field for simulating the turbulent channel flow at the Reynolds number up to ${{\mathop{\rm Re}\nolimits} _\tau } = 20000$.		
	Finally, it should be mentioned that if the flow complexity increases, e.g. by introducing the inhomogeneous effect along the streamwise or spanwise direction, or considering the rough-wall boundary, the present wall model should be extended with specific designs. These aspects warrant further investigation in future studies.
	
%
%
%
		
	\begin{acknowledgments}
		We are grateful to Prof. Xiang I. A. Yang at Pennsylvania State University and Dr. Zhi-Wen Cui at Tsinghua University for providing valuable suggestions for this research. This work was supported by the National Natural Science Foundation of China under Grant Nos. 12425206, 12272206, 92252204 and 12388101.
	\end{acknowledgments}
	
	\appendix
	
	\section{Effect of the modification layers}\label{sec:appA}

\begin{figure}
		\centering
		\begin{overpic}[width=0.48\linewidth
			]{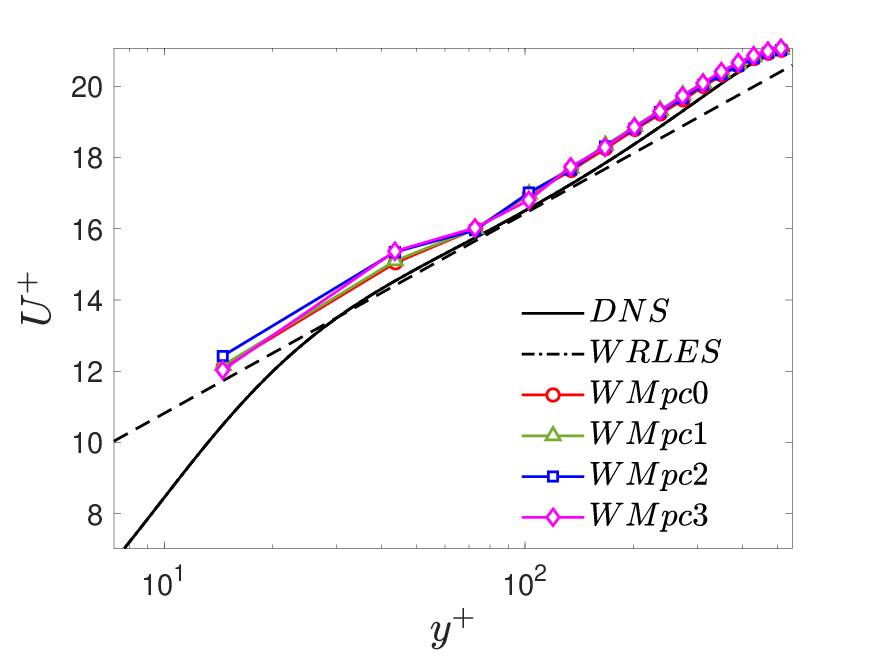}
			\put(7,73){($a$)}
		\end{overpic}
		\begin{overpic}[width=0.48\linewidth
			]{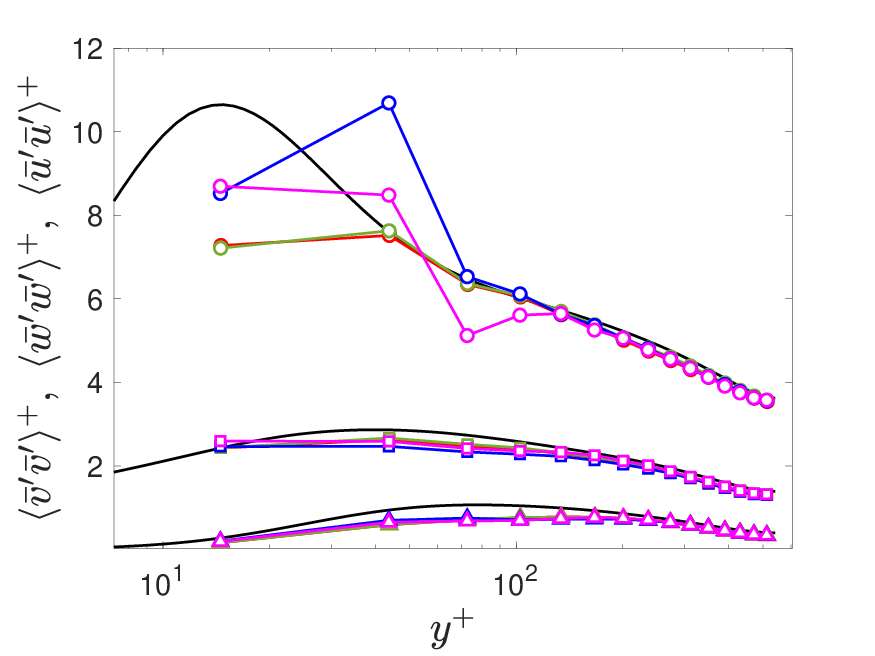}
			\put(7,73){($b$)}
		\end{overpic}\\
		\begin{overpic}[width=0.48\linewidth
			]{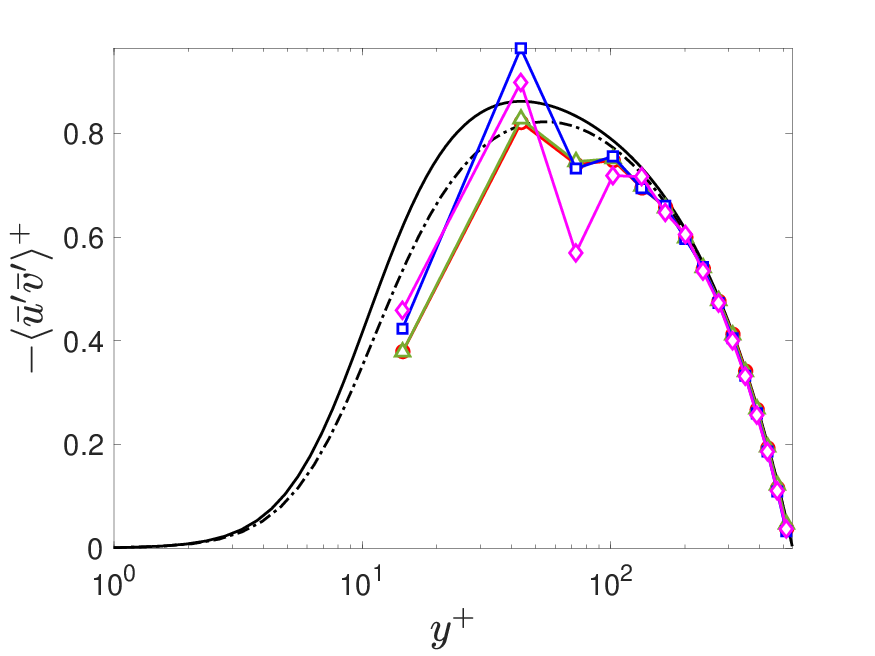}
			\put(7,73){($c$)}
		\end{overpic}
		\begin{overpic}[width=0.48\linewidth
			]{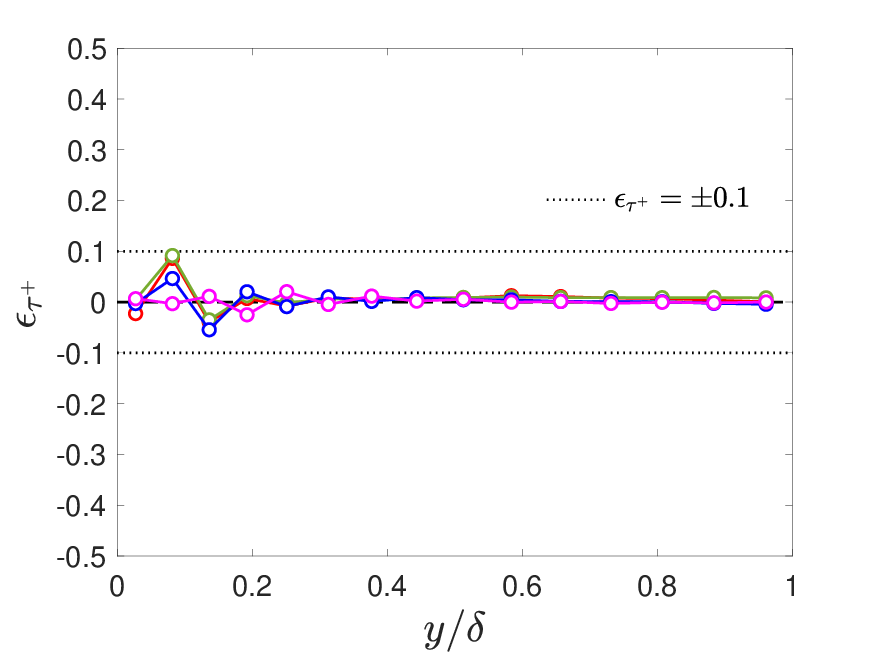}
			\put(7,73){($d$)}
		\end{overpic}\\
		\quad\\
		\captionsetup{justification=raggedright,singlelinecheck=true}
		\caption{(Colour online) Simulation results of the WMpcn ($n=0-3$) cases at ${Re}_\tau=550$: (a) profiles of mean velocity $U=\langle \bar{u} \rangle$, where the black dashed line denotes the logarithmic law, i.e. $U^+=\frac{1}{0.41}ln(y^+)+5.2$; (b) the resolved Reynolds normal stresses, where the streamwise and spanwise components are shifted upward by 3 units and 1 unit, respectively, for clarity; (c) Profiles of resolved Reynolds shear stress; (d) deviation of the total shear stress, i.e. ${\epsilon_{{\tau ^ + }}} = {\left\langle \rho {\left( {\nu  + {{\bar \nu }_{sgs}}} \right)\frac{{\partial \bar u}}{{\partial y}}} \right\rangle ^ + } - {\left\langle {\rho \bar u\;\bar v} \right\rangle ^ + } - \left( {1 - \frac{y}{\delta }} \right)$. The DNS data from \cite{lee_direct_2015} and the WRLES results (see the caption of figure \ref{fig:WMn-models}) are also plotted for comparison.}
		\label{fig:modLayer-abcd}
	\end{figure}

\begin{figure}
		\centering
		\begin{overpic}[width=0.9\linewidth
			]{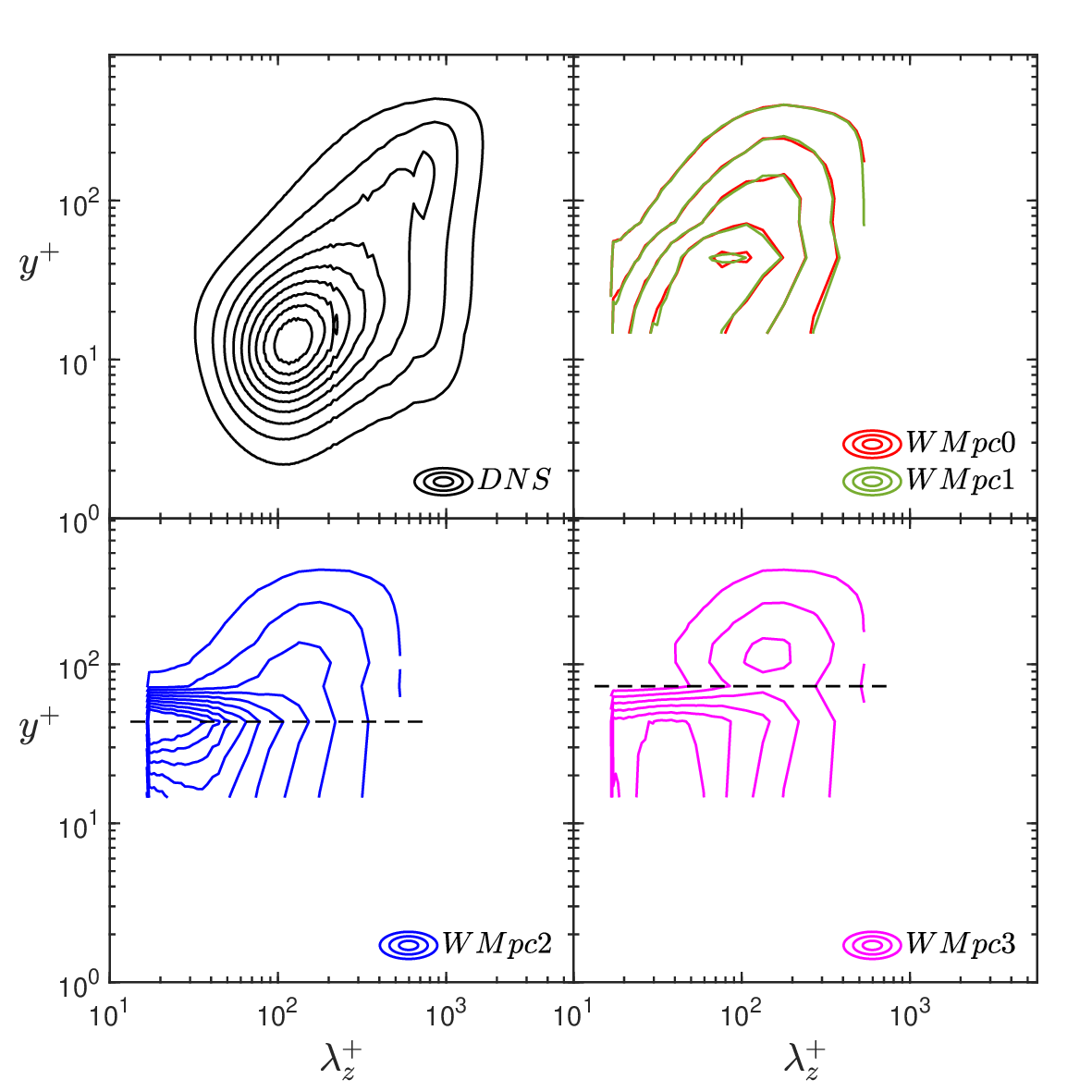}
			\put(11,92){($a$)}
			\put(54,92){($b$)}
			\put(11,49){($c$)}
			\put(54,49){($d$)}
		\end{overpic}
		\captionsetup{justification=raggedright,singlelinecheck=true}
		\caption{(Colour online) Contours of the spanwise premultiplied spectra of streamwise velocity fluctuation ($k_z\phi_{\bar{u}^\prime\bar{u}^\prime}/u_\tau^2$) for turbulent channel flow at ${Re}_\tau=550$: (a) DNS data from \cite{lee_direct_2015}; (b) WMpc0 and WMpc1 models; (c) WMpc2 model; (d) WMpc3 model. The modification zones are below the black dashed lines. Contours are from 0 to 4 with an interval of 0.4. }
		\label{fig:modLayer-spectral}			
	\end{figure}

\begin{figure}
		\centering
		\begin{overpic}[width=0.48\linewidth
			]{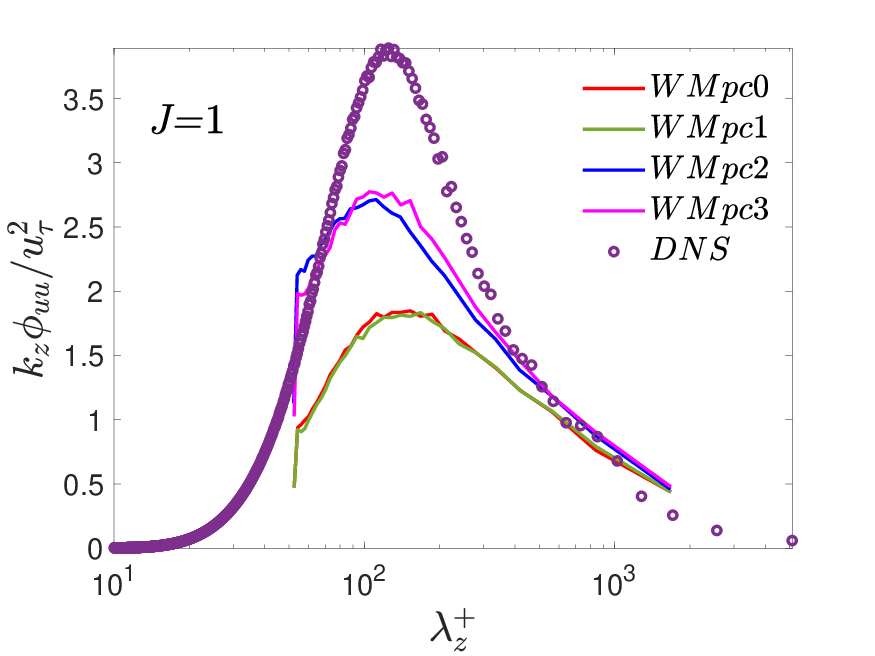}
			\put(7,73){($a$)}
		\end{overpic}
		\begin{overpic}[width=0.48\linewidth
			]{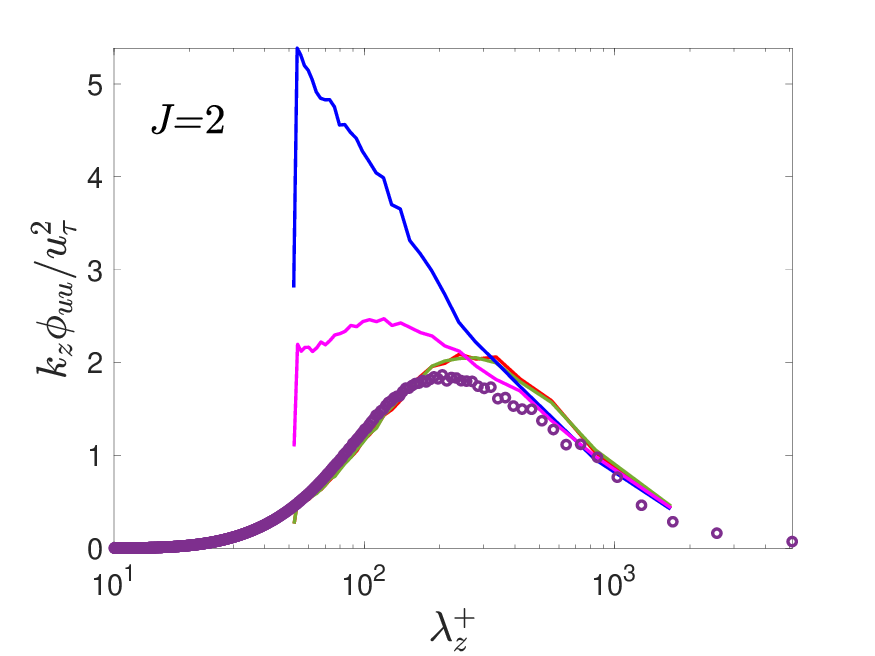}
			\put(7,73){($b$)}
		\end{overpic}\\
		\begin{overpic}[width=0.48\linewidth
			]{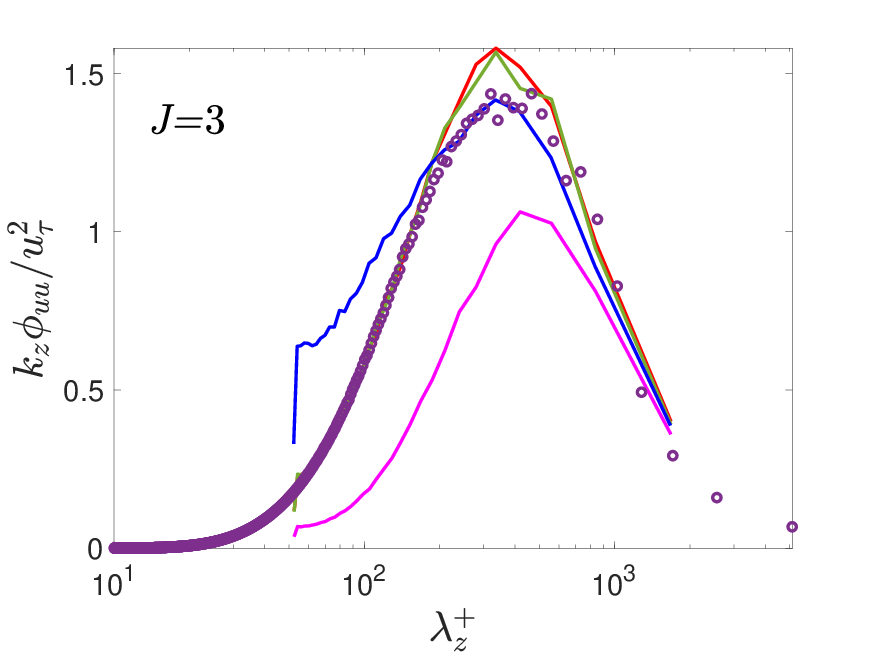}
			\put(7,73){($c$)}
		\end{overpic}
		\begin{overpic}[width=0.48\linewidth
			]{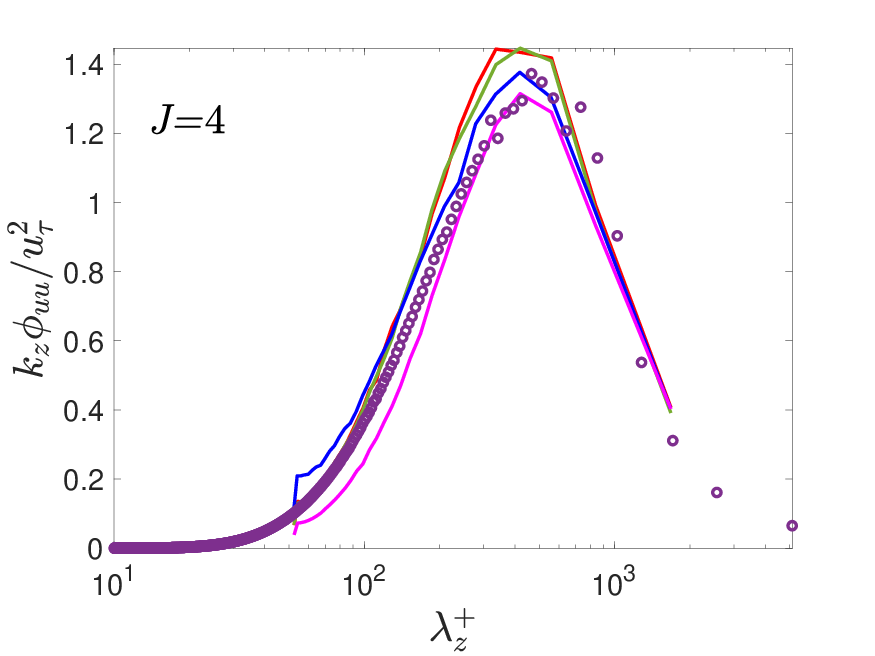}
			\put(7,73){($d$)}
		\end{overpic}\\
		\quad\\
		\captionsetup{justification=raggedright,singlelinecheck=true}
		\caption{(Colour online) Comparison of the spanwise premultiplied spectra for streamwise velocity fluctuation ($k_z\phi_{\bar{u}^\prime\bar{u}^\prime}/u_\tau^2$) from the WMpcn ($n=0-3$) models and DNS data of \cite{lee_direct_2015} at (a) the first, (b) the second, (c) the third and (d) the fourth grid, for turbulent channel flow at ${Re}_\tau=550$. The circles represent the DNS data at the same height in wall units.}
		\label{fig:modLayer-J1234}
	\end{figure}

	In this appendix, we discuss the effect of the modification layers by using the TSSC model (or the WMpcn model). The results in figure~\ref{fig:test-TSSC} display that the effect of the modification layers is weak when the grid resolution is coarse enough, i.e., $\Delta {x^ + } \mathbin{\lower.3ex\hbox{$\buildrel>\over{\smash{\scriptstyle\sim}\vphantom{_x}}$}} 200$ and $\Delta {x^ + } = \Delta {z^ + }$ or $\Delta {x^ + } = 2\Delta {z^ + }$. Here we apply the TSSC model at ${{\mathop{\rm Re}\nolimits} _\tau } = 550$ by using relatively high-resolution grids, with uniform grid along the streamwise and spanwise directions ($\mathrm{\Delta}_x^+\approx54$, $\mathrm{\Delta}_z^+\approx27$), and stretched grid along the wall-normal direction ($\mathrm{\Delta}_y^+\approx30-43$). The wall shear stress errors are given in the table~\ref{tab:modLayer}, and the low-order turbulence statistics, the two-dimensional contours and one-dimensional spanwise premultiplied spectra for streamwise velocity fluctuations $k_z^ + \phi _{\bar u^\prime\bar u^\prime}^+$ are displayed in figures ~\ref{fig:modLayer-abcd}-\ref{fig:modLayer-J1234}, respectively. It is seen that the wall shear stress error (table~\ref{tab:modLayer}), mean velocity profile (figure~\ref{fig:modLayer-abcd}a) are only slightly affected by the coverage of the modification zone. With the modification zone covering more layers, although the conservation of the total shear stress is satisfied within a wider region (see figure ~\ref{fig:modLayer-abcd}d), the profiles of the resolved Reynolds stresses become much more jagged in the near-wall region [see figure ~\ref{fig:modLayer-abcd}(b-c)]. This may be attributed to the lack of the fluctuation part of the SGS eddy viscosity calculated by the TSSC model. Hence, in the WMpc2 and WMpc3 cases, the flow structures in the modification zone are unphysical and the shapes of the corresponding spectra deviate more from the DNS data [see figure~\ref{fig:modLayer-spectral}(a,c,d)]. The spectra shown in figure ~\ref{fig:modLayer-spectral}(b), which represent the flow structures above the modification zone for the WMpc0 and WMpc1 cases, are similar with the DNS results, indicating the accurate prediction of the outer-layer flow. For the WMpc2 and WMpc3 cases, affected by the inner layer, there are some discrepancies in the spectra above the modification zone compared to the DNS results (see figure~\ref{fig:modLayer-spectral}c,d). Nevertheless, according to figure~\ref{fig:modLayer-J1234}, the influenced region is limited within one layer beyond the modification zone. For example, the magenta lines (i.e. the WMpc3 case) have a tendency approaching to the DNS results with the increase of the wall distance and nearly match the DNS data at the fourth grid.
	
	According to the above results, when the wall-parallel grid resolution is relatively high ($\Delta {x^ + },\Delta {z^ + } \mathbin{\lower.3ex\hbox{$\buildrel<\over{\smash{\scriptstyle\sim}\vphantom{_x}}$}} 50$) and the near-wall flow structures need to be considered, e.g. in the transitional, separated and reattached flows, the number of the modification layers is suggested to be less than two.
	
	
	\begin{table}
		\captionsetup{justification=raggedright,singlelinecheck=true}
		\caption{\label{tab:modLayer}
			Comparison of wall shear stress among the wall model, the LES solver, and the DNS data at ${Re}_\tau=550$ for the WMpcn ($n=0-3$) models. The grid numbers are the same for the four WMLES cases, i.e., $64(x)\times30(y)\times64(z)$ with uniform grid in the streamwise and spanwise directions ($\mathrm{\Delta}_x^+\approx54$, $\mathrm{\Delta}_z^+\approx27$), and stretched grid in the wall-normal direction ($\mathrm{\Delta}_y^+\approx30-43$).}
		\begin{ruledtabular}
			\begin{tabular}{lcccc}
				\textrm{Case}&
				\textrm{WMpc0}&
				\textrm{WMpc1}&
				\textrm{WMpc2}&
				\textrm{WMpc3} \\
				\colrule
				$\varepsilon_{LW}(\%)$ & 0.13 & 0.10 & -0.02 & -0.08 \\
				$\varepsilon_{LD}(\%)$ & -3.72 & -3.86 & -5.51 & -4.18 \\
			\end{tabular}
		\end{ruledtabular}
	\end{table}

				%
	
	\clearpage
	
	\nocite{*}
	
	\bibliography{Ref}
	
\end{document}